\documentclass{aa}  

\usepackage{graphicx}
\usepackage{gensymb}
\usepackage{verbatim}
\usepackage{siunitx}
\usepackage{amssymb}
\usepackage{amsmath}
\usepackage{setspace}
\usepackage[table,xcdraw]{xcolor}
\usepackage{tabularx}
\usepackage[normalem]{ulem} 
\usepackage{txfonts}
\usepackage[colorlinks,allcolors=blue]{hyperref}
\usepackage{placeins}

\newcommand{\starforge}{{\small STARFORGE} }
\newcommand{\starforgee}{{\small STARFORGE}}
\begin{document}

   \title{Effects of stellar feedback on cores in STARFORGE}

   \subtitle{}
   \author{K. R. Neralwar\inst{1}\thanks{Member of the International Max Planck Research School (IMPRS) for Astronomy and Astrophysics at the Universities of Bonn and Cologne.},
          D. Colombo\inst{2},
          S. Offner\inst{3},
          F. Wyrowski\inst{1},
          K. M. Menten\inst{1},
          A. Karska\inst{1,2,4},
          M Y. Grudi\'{c}\inst{5},
          S. Neupane\inst{1}
          }%

   \institute{Max-Planck-Institut f\"ur Radioastronomie, Auf dem H\"ugel 69, 53121 Bonn, Germany \\ email: \texttt{kneralwar@mpifr-bonn.mpg.de}
   \and
   Argelander-Institut f\"ur Astronomie, Auf dem H\"ugel 71, 53121 Bonn
   \and
    Department of Astronomy, The University of Texas at Austin, Austin, TX 78712, USA
   \and
   Institute of Astronomy, Faculty of Physics,
Astronomy and Informatics, Nicolaus Copernicus University, ul. Grudziądzka 5, 87-100 Toruń,
Poland
    \and
    Carnegie Observatories, 813 Santa Barbara St, Pasadena, CA 91101, USA
}

  \date{Received XXX; accepted XXX}

  \abstract 
  {Stars form in dense cores within molecular clouds and newly formed stars influence their natal environments. How stellar feedback impacts core properties and evolution is subject to extensive investigation. 
  We performed a hierarchical clustering (dendrogram) analysis of a \starforge simulation modelling a giant molecular cloud to identify gas overdensities (cores) and study changes in their radius, mass, velocity dispersion, and virial parameter with respect to stellar feedback.
  We binned these cores on the basis of the fraction of gas affected by protostellar outflows, stellar winds, and supernovae and analysed the property distributions for each feedback bin. 
  We find that cores that experience more feedback influence are smaller. 
  Feedback notably enhances the velocity dispersion and virial parameter of the cores, more so than it reduces their 
  radius. 
  This is also evident in the linewidth-size relation, where cores in higher feedback bins exhibit higher velocities than their similarly sized pristine counterparts. 
  We conclude that stellar feedback mechanisms, which impart momentum to the molecular cloud, simultaneously compress and disperse the dense molecular gas.
  }
  
   \titlerunning{Feedback effects on cores in \starforge}
   \authorrunning{K. R. Neralwar et al.}
   \maketitle


\section{Introduction}\label{sec: intro}

Stellar feedback is one of the most important processes that regulate star formation in a galaxy \citep{krumholz2014, agertz2015, gatto2017, peters2017MNRAS.466.3293P,kruijssen2019, guszejnov2022}. It can have a positive as well as a negative impact on star formation rates; i.e. it can trigger the formation of a new star by compressing gas or limit star formation by destroying the natal cloud. Simulations show that in the absence of stellar feedback, the gas cools rapidly, forming dense structures, which does not lead to the multiphase interstellar medium (ISM) seen in a multitude of observations \citep{schuller2009A&A...504..415S, colombo2019, duarte_cabral2021}.

Stellar feedback mechanisms include protostellar outflows, stellar winds, radiation pressure, photoionisation, and supernova explosions \citep{krumholz2014, girischidis2020SSRv..216...68G}. These different mechanisms operate at different spatial scales and are strongly linked to the evolutionary stages of stars \citep{hopkins2012}.
Feedback from massive stars provides an endpoint for star formation \citep{geen2015MNRAS.448.3248G, grudic2022}. 

Protostars expel gas in the form of bipolar jets \citep{fendt2002A&A...395.1045F, wu2004A&A...426..503W, bally2016}. Expelled material entrains molecular gas from the surrounding core and accelerates it to high velocities, thus creating molecular outflows. 
These outflows have complex morphologies and clumpy structures \citep{arce2001ApJ...554..132A} that evolve over time \citep{lee2002ApJ...576..294L,Offner2011}. They inject momentum and energy into the ISM over scales ranging from a few AU to tens of parsecs \citep{frank2014prpl.conf..451F, bally2016}. Protostellar outflows reduce stellar mass by ejecting accreting gas. They also remove excess angular momentum and facilitate reduced accretion rates during the protostellar phase of 
young stellar objects \citep{arce2007prpl.conf..245A, federrath2014ApJ...790..128F, bally2016,offner2017, guszejnov2021MNRAS.502.3646G, guszejnov2022}. Outflows unbind the natal core gas, which can reduce the star formation efficiency (SFE) of the dense gas \citep{offner2017}.

Stellar winds are generated by the ejection of matter from the stellar surface due to the impact of radiation pressure on the gas in the stellar atmosphere \citep{arce2011ApJ...742..105A, pabst2020A&A...639A...2P, geen2023}. Winds from massive stars profoundly affect the ISM and lead to complex molecular gas structures \citep{zinnecker2007ARA&A..45..481Z, xu2020}.
They often manifest as shells/bubbles around individual stars that are often associated with HII regions \citep{deharveng2010A&A...523A...6D, schneider2020PASP..132j4301S, kirsanova2023IAUS..362..268K}. 
Stellar winds do not 
strongly affect the global cloud evolution but aid in preventing runaway accretion of massive stars \citep{guszejnov2022}.

Supernovae mark the endpoints of stellar evolution for massive stars and are a key ingredient in the study of interstellar gas \citep{mckee1977ApJ...218..148M, dubner2015A&ARv..23....3D}. They act on scales up to $\sim 100$ pc and inject a large amount of energy ($\sim10^{51}$ ergs) that drives turbulence into the ISM, 
and they regulate the chemical composition of the ISM by producing metals and dust \citep{wesson2021ApJ...923..148W, kirchschlager2024MNRAS.tmp..359K}. These explosions inject material and drive shocks into the ISM, and create supernova remnants \citep[SNRs; ][]{dubner2015A&ARv..23....3D, dokara2023A&A...671A.145D}. 

The various stellar feedback mechanisms play an important role in shaping the internal structure of giant molecular clouds (GMC) and constraining their lifetimes \citep{williams1997ApJ...476..166W, hopkins2012, chevance2023}. The internal structure of MCs is often hierarchically classified according to their size and mass. GMCs are the largest molecular gas structures in the galaxy and they form the denser molecular ISM in the Galaxy \citep{blitz1993prpl.conf..125B, rosolowsky2021MNRAS.502.1218R, chevance2023}. Cores are gas structures with sizes around 0.1 pc or less \citep{ballesteros-paredes2020}. Some dense cores accrete gas and become massive enough to collapse under self-gravity \citep{offner2022MNRAS.517..885O}, leading to the formation of stars.

New observational capabilities, in particular the high angular resolution, combined with high sensitivity afforded by the Atacama Large Millimeter/submillimeter Array (ALMA) have led to observations of large samples of cores in the Galaxy \citep{kramer2023arXiv231001044K, nony2023A&A...674A..75N, olmi2023MNRAS.518.1917O}, but how stellar feedback affects them is still not well understood. 
Simulations allow us to probe sub-core spatial scales and investigate how feedback operates { on cores and shapes their properties} \citep{offner2017}.

In this study we examine the effect of outflows, winds, and supernovae on cores based on a \starforge simulation of a GMC that resolves the formation of individual stars and follows the star formation process until it is self-consistently halted by stellar feedback. We explore how different feedback mechanisms influence the physical properties of cores, i.e. radius, mass, velocity dispersion, and virial parameter.
Section \ref{sec: data} describes the \starforge simulation that we analyse. We performed a dendrogram analysis on the simulation snapshots to obtain the "cores". We discuss it along with the methods for obtaining the core properties in Sect. \ref{sec: methods}. We present and discuss the effects of the three feedback mechanisms on the core properties in Sect. \ref{sec: results}, where we also address the well-known scaling relations from the star formation literature. Finally, we summarise our findings in Sect. \ref{sec: summary}.

\section{\starforge}\label{sec: data}

The \starforgee\footnote{\url{https://www.starforge.space}} simulations are 3D radiation magnetohydrodynamic (MHD) simulations that follow the life of GMCs with a typical maximum spatial resolution of ($\sim$ 10 AU). The framework is implemented in the {\small GIZMO} code \citep{gizmo}, which employs a Lagrangian meshless finite-mass (MFM) method to solve the MHD equations \citep{hopkins2016MNRAS.455...51H}. The simulations follow the formation, accretion, evolution, and dynamics of individual stars in a giant molecular cloud and include all feedback mechanisms: jets, radiation, stellar winds, and supernovae. A detailed description of the numerical methods and relevant tests can be found in \cite{grudic2021}.

Jets, stellar winds, and supernovae are introduced as mass injection events in the simulations. The mass is injected from star particles into the simulation domain by two processes: local injection and cell spawning. Local injection refers to the weighted distribution of fluxes (mass, momentum, energy) to neighbouring cells. This method is used for photon injection and to propagate stellar winds when the free-expansion radius is unresolved. Cell-spawning refers to the process of creating new finite-mass gas cells at a certain rate, directly resolving the flow of feedback around the star. This is performed for protostellar jets and for winds when the free-expansion radius can be resolved at the given mass resolution.

In \starforgee, protostellar outflows are modelled based on the prescription in \citet{cunningham2011ApJ...740..107C}, using three parameters: (i) the fraction of mass accreted by the disc that is diverted to the jet (f$_w = 30\%$), (ii) the fraction of Keplerian velocity (v$_{jet}$) at the protostellar radius and (iii) the collimation angle $\theta_0$, which sets the angular distribution of the injected jet and outflow momentum. Main-sequence stars with mass M > 2 M$_\odot$ generate stellar winds with a mass loss rate and wind velocity given by the equations. 44-45 in \citet{grudic2021}. All stars with mass M > 8 M$_\odot$ are modelled to end as supernova at the end of their lifetimes (following Equation 47 in \citet{grudic2021}).
A detailed description of how the feedback mechanisms are modelled in \starforge is given in Section 4 in \cite{grudic2021}.

In this work, we analyse the fiducial \texttt{M2e4} simulation described in Table 1 of \cite{guszejnov2022}. This particular suite of simulations follows the evolution of an { initially spherical} cloud of 2 $\times$ $10^4$ M$_\odot$ with an initial radius ($R_{cloud}$) of 10 pc, discretised with a mass resolution of $\delta m=10^{-3}M_\odot$. The simulation contains 454 snapshots with an equal spacing of 24.7 kyr.
The initial uniform magnetic field is in the $\hat{z}$ direction and corresponds to a mass-to-magnetic flux ratio ($\mu$) of 4.2. The cloud is initialised with a Gaussian random velocity field scaled for an initial virial parameter $\alpha = $ 2. The first stars form and launch protostellar outflows at 0.8 Myr. The first massive stars join the main-sequence star and begin producing stellar winds at 3.6 Myr. It is important to note that winds and radiation halt star formation before the first supernova occurs at 9.8 Myr \citep{grudic2022,guszejnov2022}.

The \starforge simulation tracks gas originating from jets (protostellar outflows), stellar winds, and supernovae using tracer fields. 
We refer to the mass of the feedback material contained in a cell relative to the mass of the non-feedback gas in the same cell as the \lq\lq feedback fraction.'' We define the gas associated with a particular feedback process as \lq\lq feedback gas.''
We use the outflow, wind, and supernova feedback fractions as a measure of how much a particular type of feedback has affected a core, e.g., the cores in low feedback bins have less influence from feedback mechanisms compared to those in high feedback bins.
However, it is important to note that the feedback fractions track only the mass of the gas launched by the different feedback mechanisms. Thus, we miss regions, such as shocks produced by photoionisation, that are influenced by radiative feedback, which does not inject mass. \starforge implements feedback by injecting new cells \citep[see][for details]{grudic2021}.
The different feedback fractions are not directly comparable to each other, because the mass fraction does not fully encapsulate the  relative significance of the feedback as it doesn't reflect velocity information.
Thus, 
we bin the cores based on these feedback fractions (Sect. \ref{sec: three feedback bins}). These bins indicate how much a feedback has influenced a core and are therefore a better tool to compare the influence of different feedback mechanisms on cores. For example, cores in high outflow bins are the ones most affected by outflows and cores in high-wind bins are the ones most affected by winds. We therefore compare these cores most affected and least affected by different feedback mechanisms using feedback bins.

\section{Methodology}\label{sec: methods}

\subsection{Dendrograms and cores}\label{sec: core selection criteria}
We use a dendrogram analysis \citep[e.g.,][]{rosolowsky2008ApJ...679.1338R} to identify the cores. Dendrograms are widely used to decompose molecular emission distributions into discrete structures such as clouds, clumps, and cores in simulations \citep[][and references therein]{smullen2020MNRAS.497.4517S, offner2022MNRAS.517..885O} and observations \citep{seo2015ApJ...805..185S, friesen2016ApJ...833..204F, keown2017ApJ...850....3K, colombo2019, duarte_cabral2021, oneil2021AAS...23711203O}. 
In our case, the dendrogram is an abstraction of the hierarchical structure of the simulated molecular cloud gas density. In essence, a dendrogram describes how isosurfaces (or three-dimensional contours) are nested within each other in the position-position-position (PPP) datacube. Dendrograms are composed of three classes of structures: trunks, branches, and leaves. Leaves are three-dimensional structures formed by single local maxima in the molecular gas distribution.  The local maxima are identified through a nearest neighbor search.  Here, we do not flatten the simulation data to a grid but instead use the {\it SciPy} {\sc kdtree} method within the dendrogram algorithm to identify the six closest neighbors to each cell.
This allows us to construct the dendrogram using the native simulation resolution.
We define the cores as the leaves of the dendrogram structure. 

We adopt dendrogram parameters informed by the properties of typical observed cores, which are estimated to have densities $\geq 10^4$ and sizes of $\sim 0.1$ pc. 
In the implementation of the dendrogram algorithm used here ({\sc astrodendro}), leaves must contain more than a minimum number of volumetric pixels\footnote{gas cells in \starforge} (\textit{min\_npix}). 
In addition, \textit{min\_value} is the minimum value considered in the dataset when constructing the dendrogram, and \textit{min\_delta} sets the significance level at which a leaf is considered to be an independent structure \footnote{A detailed description of these dendrogram variables can be found here : \url{https://dendrograms.readthedocs.io/en/stable/using.html}.}.
We set \textit{min\_npix} to 100, \textit{min\_value} to an H$_2$ number density of $10^4\,\rm{cm}^{-3}$ and \textit{min\_delta} to $10^{4}\, \rm{cm}^{-3}$, following \cite{offner2022MNRAS.517..885O}.  
The \textit{min\_npix} = 100 corresponds to a minimum leaf mass of 0.1 $M_\odot$, which is comparable to the completeness limits of the core surveys \citep[e.g.,][]{sokol2019MNRAS.483..407S}.
The chosen peak threshold filters out low-density structures that would otherwise not be identified as cores in the observational data \citep{chen2019}.  While {\it Gizmo} estimates the fraction of mass in each cell that is ionised, neutral, and molecular,  we make no explicit assumption about the phase or temperature of the gas at these densities. All gas in a given cell that satisfies the density threshold is included in the analysis.

We aim to identify the causation between the core properties and the presence of feedback, and not just the correlation. For example, if a structure with a significant amount of directly launched outflow material is included in the core identification, such a core will naturally tend to have a higher velocity dispersion, even if most of the core is largely unaffected. We instead aim to explore the properties of cores affected by the feedback as seen through molecular line observations, e.g., by excluding hot outflow and wind material that is traced by the atomic and ionised gas. Our cores mostly contain gas that is traced by various molecular lines such as $^{12}$CO, $^{13}$CO, C$^{18}$O and NH$_3$, where the relatively dense and cold ISM varies in feedback influence. 
To exclude raw (recently injected) feedback material, we mask any pure feedback cells that, by construction, have masses below the fiducial gas mass resolution of $10^{-3}$M$_\odot$ (see \citealt{grudic2021}). Some feedback gas may otherwise be included by the above density threshold criterion, since the outflow material launched by the protostar subgrid model is often directly injected into the relatively dense core environment.
Moreover, if we include these pure feedback cells when constructing leaves, we retrieve some cores with a mass less than 0.1 M$_\odot$, below typical observational completeness limits. 
Our dendrogram parameter $min\_npix$ acts as a minimum mass threshold of $min\_npix \times \Delta m$, given $\Delta m \geq 10^{-3}$M$_\odot$.
Consequently, our dendrogram structures include only well-mixed feedback material, which has already interacted (merged) with the surrounding envelope, similar to observational expectations.

\subsection{Core properties}

We used the properties of simulation cells grouped as a leaf by dendrogram analysis to obtain four independent properties of the cores\footnote{We use the terms leaf and core interchangeably throughout the paper. Hereafter, we refer to cells that belong to a particular leaf of the dendrogram as \lq\lq leaf cells.''}: radius, velocity dispersion, mass and virial parameter.

We define the core radius as 
\begin{equation}\label{eqn: core radius}
    r = \sqrt{\frac{5}{2}  \frac{\sum r_1^2  m_{\rm{cell}}}{\sum m_{\rm{cell}}}}
\end{equation}
where,  $r_1 = \sqrt{dx^2 + dy^2 + dz^2}$ and $d_i = i_{\rm{peak}}$\footnote{The leaf peak cell ($i_{\rm{peak}}$) refers to the cell in the leaf with the maximum density as determined by the dendrogram.}
- $i_{\rm{cell}}$ (i = x,y,z). The radius (Eqn. \ref{eqn: core radius}) is obtained by assuming the cores to be spherical and using the moment of inertia relation for a sphere. 
We define the 1D velocity dispersion for each leaf as 
\begin{equation} \label{eqn: core 1d dispersion}
    \sigma_{1d} =  \sqrt{\frac{\sum (v_{i}^{\rm{cell}} - \bar{v}_{i}^{\rm{cell}})  m_{\rm{cell}}}{3 \sum m_{\rm{cell}}}}
\end{equation}
where $v_{i}^{\rm{cell}}$ represent the cell velocities in the x,y,z directions.
We define the core mass as the sum of the molecular gas mass of the leaf cells, where we use the neutral hydrogen abundance ($f_{\rm{neutral}}$) and the molecular mass fraction ($f_{\rm{mol}}$) to determine the amount of molecular material in each cell:

\begin{equation} \label{eqn: core mass}
    m_{\rm{core}} = \sum m_{\rm{cell}} \times f_{\rm{neutral}} \times f_{\rm{mol}}.
\end{equation}
{ This mass is comparable to the total gas mass in the cells, since most of the identified cores are composed of predominantly cold and molecular gas.}

The core virial parameter $\alpha_{vir}$
is the ratio of the kinetic energy to the potential energy.
We define the core kinetic energy as 
\begin{equation}\label{core KE}
    K.E. =  \sum\frac{m_{\rm{cell}}(\sigma_x^2+\sigma_y^2+\sigma_z^2)}{2}.
\end{equation}
The potential energy is obtained using the gravitational potential calculation function \footnote{\url{https://github.com/mikegrudic/pytreegrav/blob/aa88a552206d22c194d0cdc3399c4b5d256c0990/src/pytreegrav/frontend.py\#L68}} \citep{pytreegrav}.

We use the feedback fractions of the cells, $f_{o}^{\rm{cell}}$, $f_{w}^{\rm{cell}}$ and $f_{s}^{\rm{cell}}$, to calculate the gas mass affected by each type of feedback for each cell (m$_{\rm{cell, o}}$, m$_{\rm{cell, w}}$ and m$_{\rm{cell, s}}$). The sum of these masses for the leaf cells gives the total feedback mass for the core. The difference between the core mass and the mass affected by different feedback gives the non-feedback core mass.
The ratio of the core feedback mass to the non-feedback mass gives the respective feedback fractions ($f_{\rm{o}}$, $f_{\rm{w}}$ and $f_{\rm{s}}$) for the cores.
\begin{equation}
    f_{x} = \sum  \frac{m_{\rm cell, x}}{m_{\rm cell} - m_{\rm{cell, y}} - m_{\rm{cell, z}}}
\end{equation}

where x, y, z = \{o,w,s\} corresponds to outflows, winds, and supernovae, respectively.

\subsection{Outflow, wind and supernova bins}\label{sec: three feedback bins}

Based on the outflow, wind, and supernova feedback fractions, $f_{\rm{o}}$, $f_{\rm{w}}$, and $f_{\rm{s}}$, we bin the cores and analyse their properties. The `no feedback' bin contains pristine cores that have zero feedback ($f_{\rm{o}}$= 0,  $f_{\rm{w}}$ = 0 and $f_{\rm{s}}$ = 0). This bin remains the same for outflow, wind, supernova, and global feedback. The other three bins are based on the percentile cuts of the respective feedback fractions (Table \ref{Table: Feedback fractions}). The `low' feedback bin contains cores containing 0--25$\%$ percentile feedback gas, the `moderate' feedback  bins contain cores with 25--75 $\%$ percentile feedback gas and the `high' feedback bins contain cores with $>$ 75 $\%$ percentile feedback gas. These bins differ for outflows, winds and supernovae as they are calculated using the percentile cuts on $f_{\rm{o}}$, $f_{\rm{w}}$ and $f_{\rm{s}}$, respectively. 
We place an additional constraint on the cores in each bin to ensure that the respective feedback is the most prominent. For example, the low-outflow bin contains only cores with $f_{\rm{w}}$ = $f_{\rm{s}}$ = 0. The moderate outflow bin has cores with $f_{\rm{w}}$ < 25\% and $f_{\rm{s}}$ < 25\%. The high outflow bin contains only cores that have  $f_{\rm{w}}$ < 75\% and $f_{\rm{s}}$ < 75\%. The wind and supernova bins follow the same criterion; e.g., the low-wind bin contains only cores with $f_{\rm{o}}$ = $f_{\rm{s}}$ = 0. This definition ensures that trends in the properties associated with a particular bin are likely the result of the respective feedback. For example, since the high-outflow bin does not have any cores with high winds and high supernovae, we can assume that the general trends in the property distributions (e.g. high velocity dispersion) of cores in this bins are a result of outflows.

\begin{table}[]
\label{Table: Feedback fractions}
\caption{The 25$^{th}$ and 75$^{th}$ percentile values of the feedback fraction distributions.}
\centering
\begin{tabular}{lcc}
\hline
Feedback  & 25 \%p$^{th}$ & 75 \%p$^{th}$ \\ \hline
Outflow   & 1.1e-03        & 2.7e-02        \\
Wind      & 7.1e-10       & 3.8e-06       \\
Supernova & 1.1e-11       & 2.6e-06        \\   \hline
\end{tabular}
\tablefoot{ We use these values to define the feedback bins in Sect. \ref{sec: three feedback bins}.}
\end{table}

\subsection{Global feedback bins}\label{sec: global feedback bins}

We define global feedback bins to study the combined effects of the three feedbacks. This ensures that our results can be compared with observational data, for which it is not possible to discriminate between the effects of the different mechanisms. The `no global' bin is identical to the `no feedback' bin. The other three bins are based on combinations of the outflow, wind, and supernova bins. The `low global' bin contains cores that belong to at least two `low feedback' bins (e.g. low outflow and low wind), with the third feedback (supernova, in this case) being empty. For example, all the cores that have 0 < $f_{\rm{o}}$ < 25\%, 0 < $f_{\rm{w}}$ < 25\% and $f_{\rm{s}}$ = 0 belong to this bin. Similarly, cores with low wind, low supernova, and no outflows also belong to the low global bin. The cores in the `moderate global' bins must belong to at least two moderate feedback bins, the third feedback being < 25\%. The `high global' bin contains cores that are in at least two high bins, the third feedback being < 75\%. The binning helps us combine the cores most affected by feedback and least affected by feedback.

\begin{table}[]
\caption{The number of cores in each feedback bin.}
\label{tab: cores in bins}
\centering
\begin{tabular}{llll}
\hline
Feedback  & Low   & Moderate & High  \\ \hline
Global    & 16994 & 83506    & 46358 \\
Outflow   & 65752 & 75297    & 41751 \\
Wind      & 0     & 5326     & 15149 \\
Supernova & 0     & 1        & 779   \\ \hline
\end{tabular}
\tablefoot{We obtain 391261 cores through the dendrogram analysis; of which, 388958 are distributed in the various feedback bins. The no-feedback bin contains 38045 cores.}
\end{table}

\begin{figure}
    \centering
    \includegraphics[width = 0.45\textwidth, keepaspectratio]{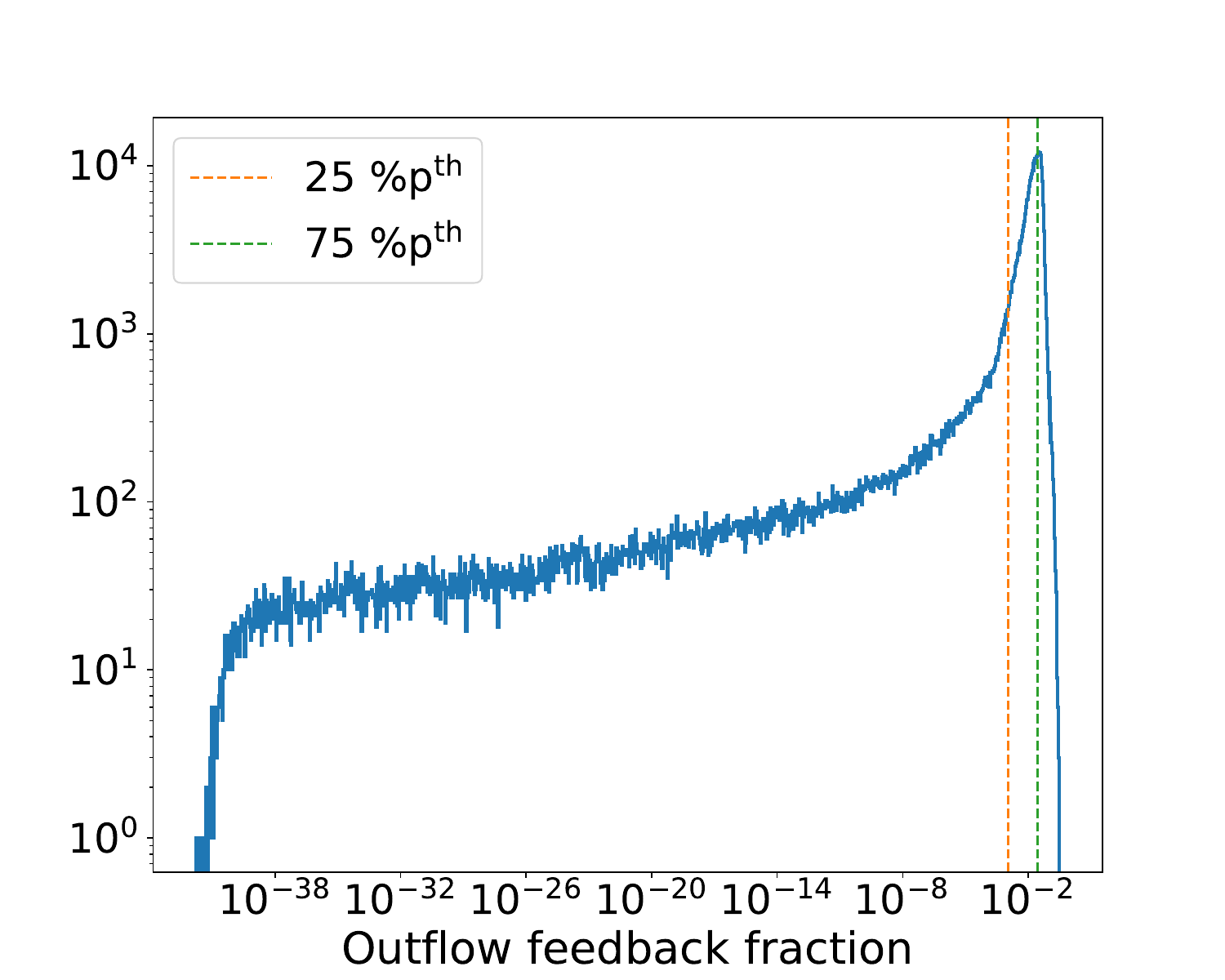}
    \includegraphics[width = 0.45\textwidth, keepaspectratio]{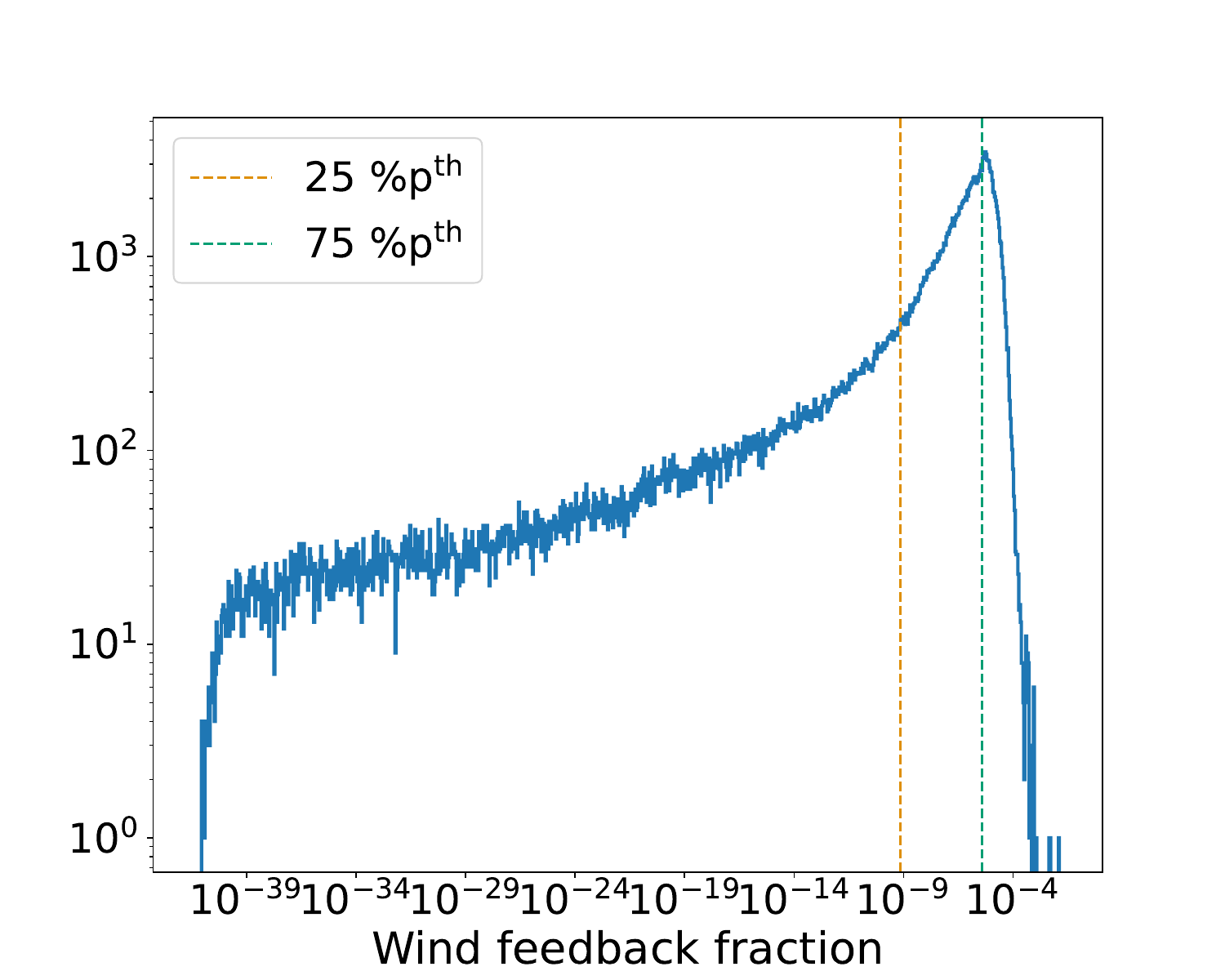}
    \includegraphics[width = 0.45\textwidth, keepaspectratio]{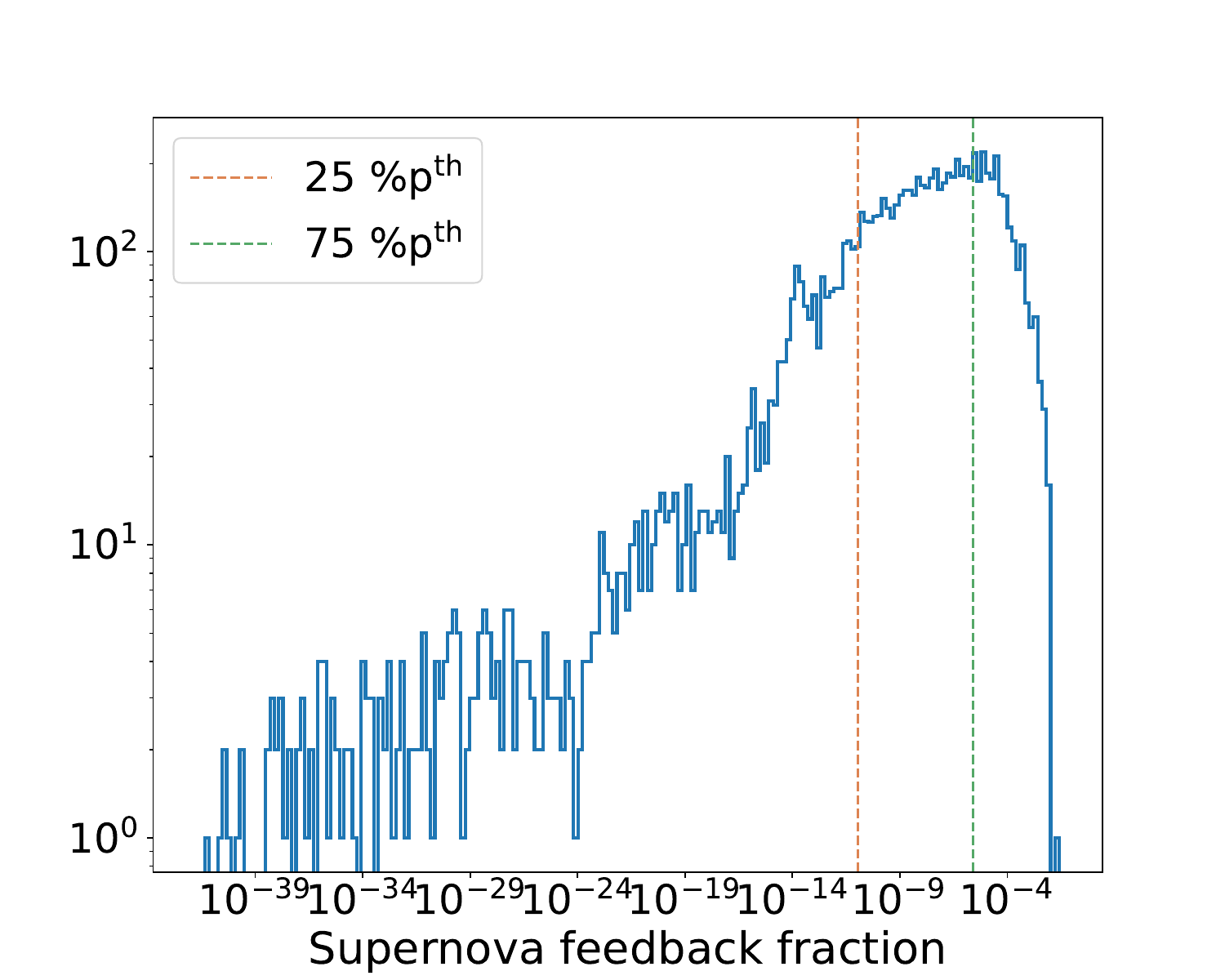}
    \caption{Histogram of outflow (top), wind (middle) and supernova (bottom) feedback fractions. We mark the 25th percentile and the 75th percentile limits for the feedback fractions.}
    \label{fig: m2e4a2 outflow hist}
\end{figure}

\section{Results}\label{sec: results}

\begin{figure*}
    \centering
    \includegraphics[width = 0.45\textwidth, keepaspectratio]{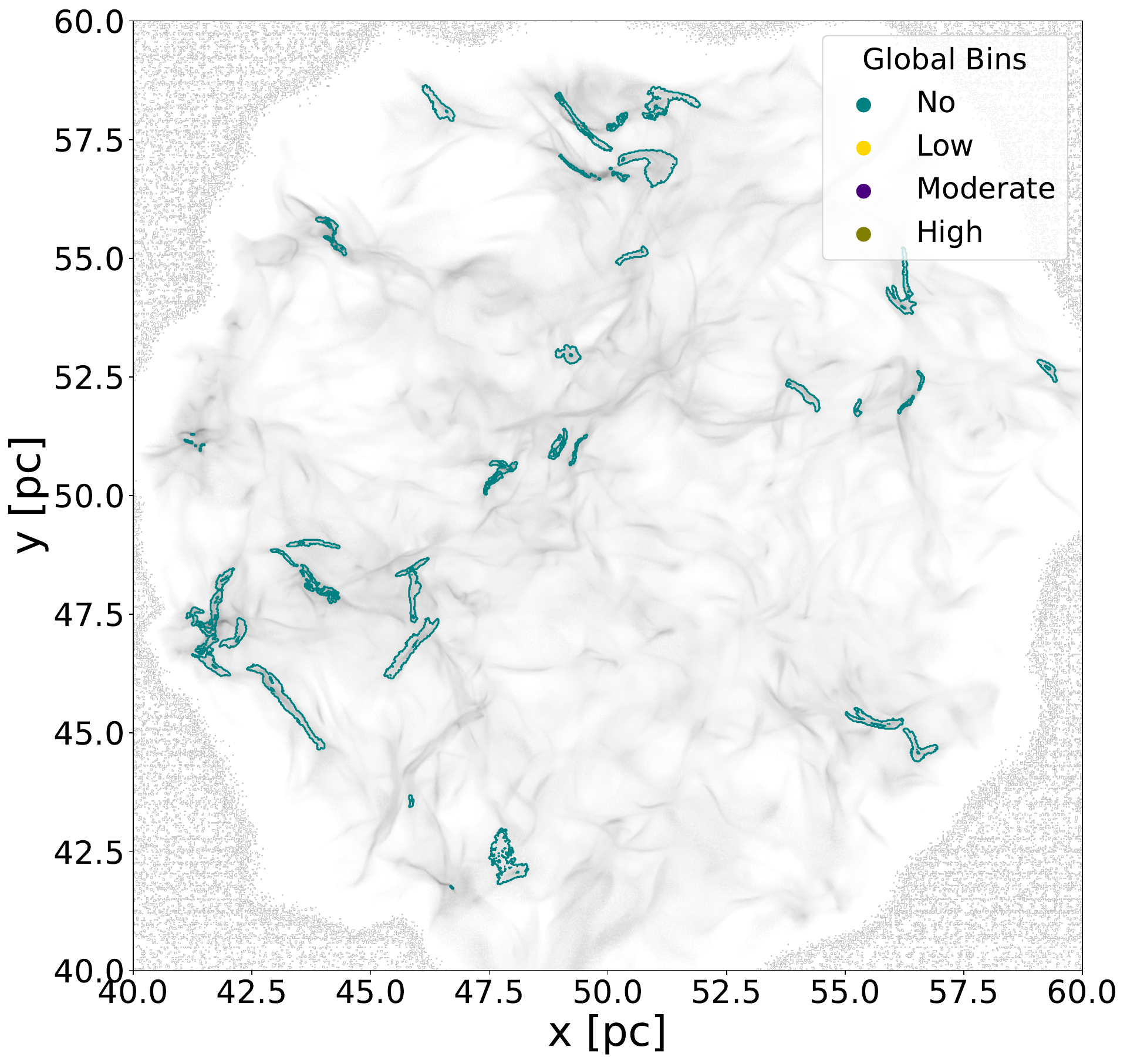}
    \includegraphics[width = 0.45\textwidth, keepaspectratio]{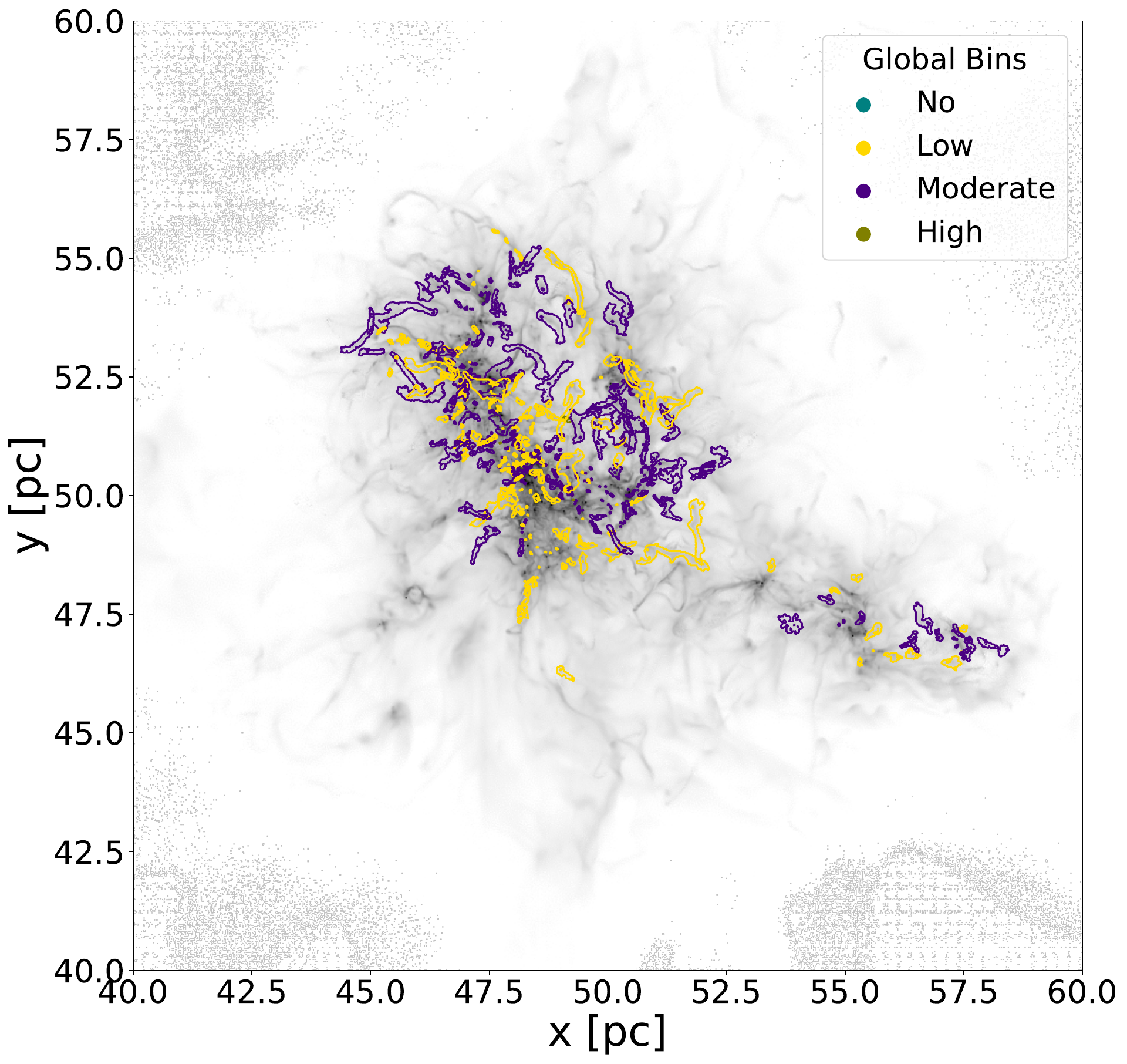}
    \includegraphics[width = 0.45\textwidth, keepaspectratio]{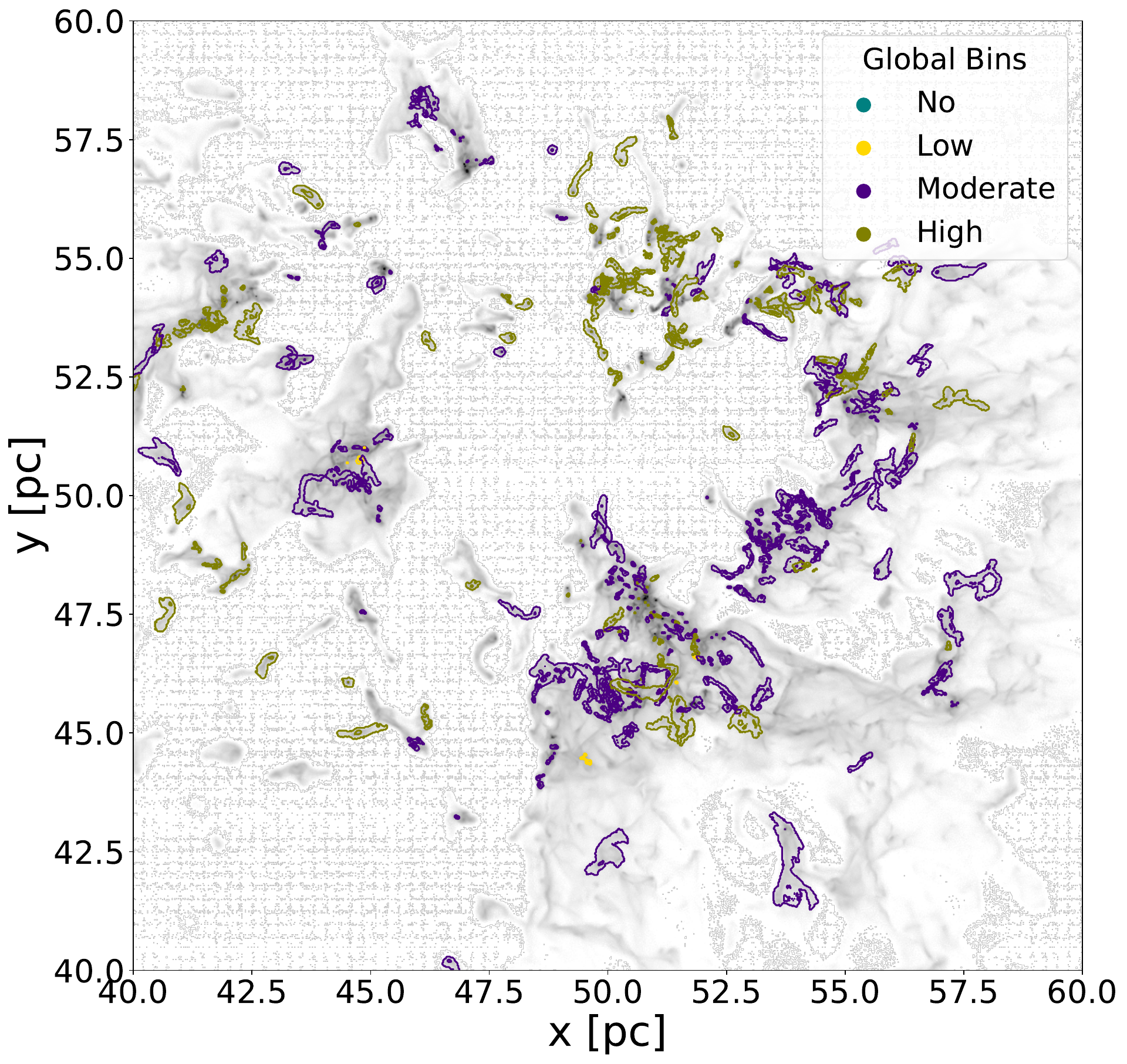}
    \includegraphics[width = 0.45\textwidth, keepaspectratio]{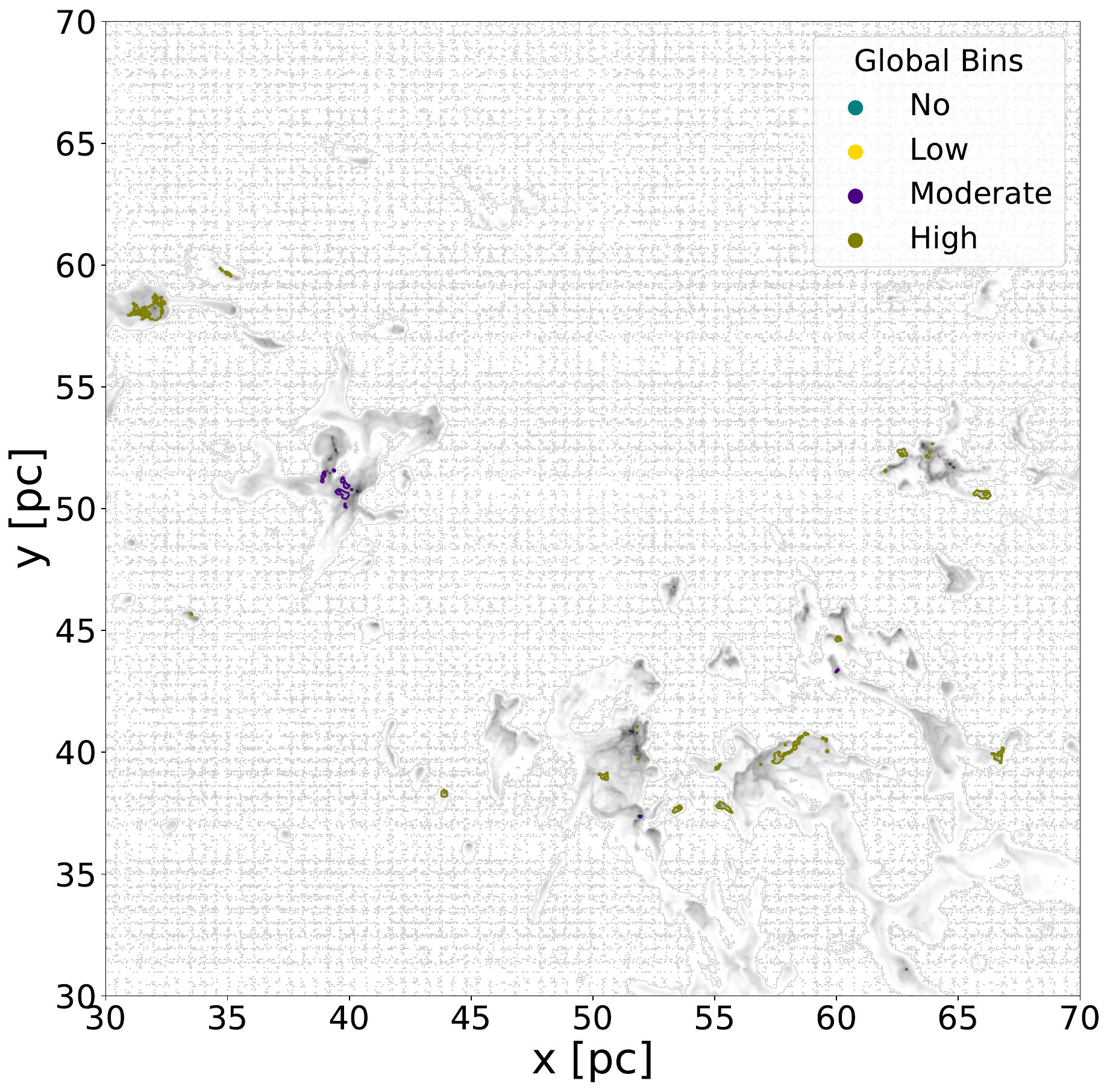}
    \caption{Visual representation of cores in four different snapshots that correspond to  0.4 Myr (top left), 4.9 Myr (top right), 7.5 Myr (bottom left), 9.9 Myr (bottom right) after the start of the simulation. The grey background shows the GMC (H$_2$ density) projected along the z axis. The contours represent the cores in different global feedback bins (Sec. \ref{sec: global feedback bins}).}
    \label{fig: core vis plots}
\end{figure*}

\begin{figure*}
    \centering
    \includegraphics[width = \textwidth, keepaspectratio]{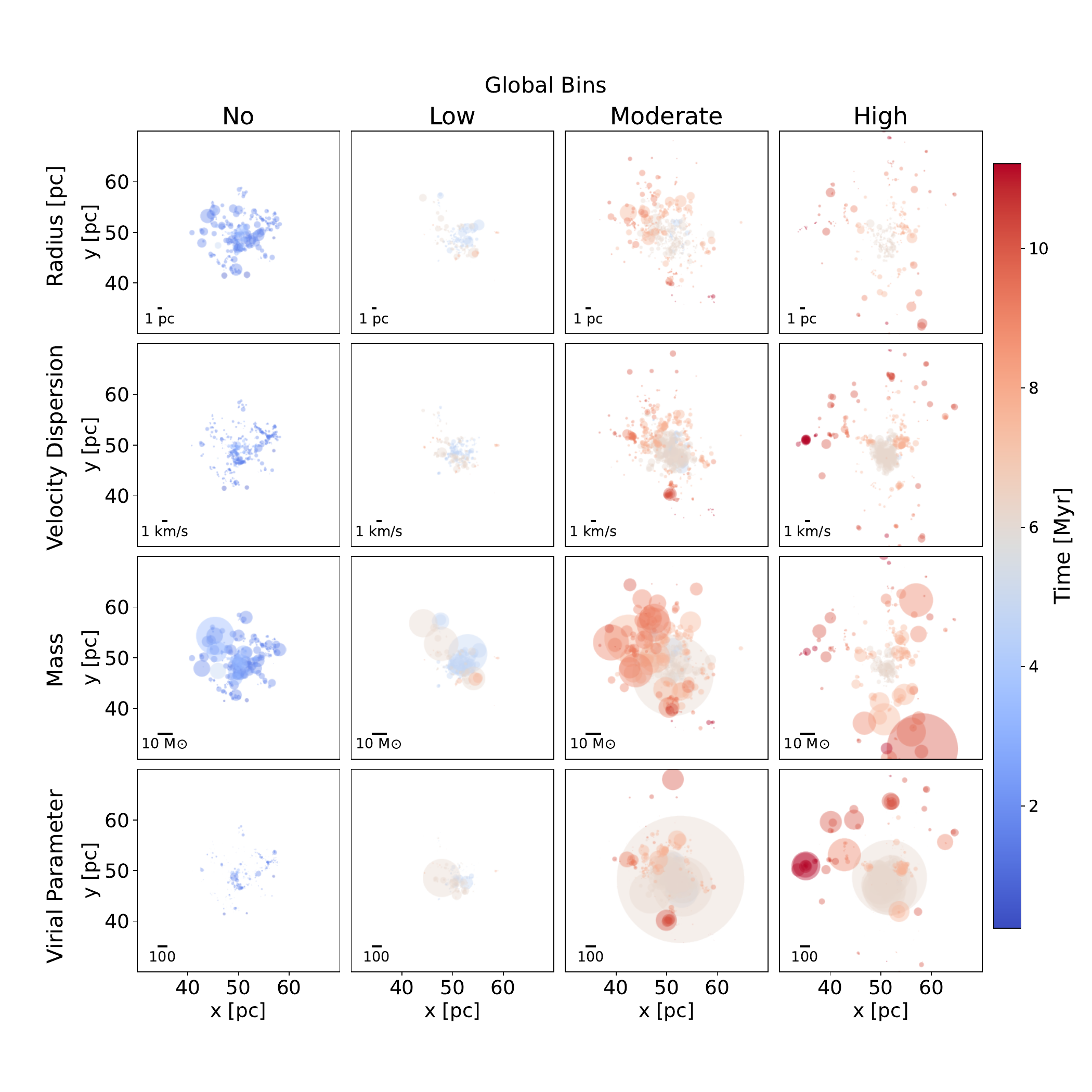}
    \caption{Distribution of cores in different snapshots along with their properties in various global feedback bins. The positions of the cores represent their locations in the simulation box. The plot assumes that the cores are circular and scales the core properties (Sect. \ref{sec: methods}) with the size of these circles. The first row visualises the radius of the core, and the second and third rows present the velocity dispersion and mass of the cores, respectively. The last row shows the virial parameter. The colours represent the time elapsed in Myr after the start of the simulation.}
    \label{fig: core feedback prop}
\end{figure*}

Figure \ref{fig: core vis plots} shows the structure of the cores as contours of the dendrogram leaves (cores) plotted on top of the GMC from which they were derived, for four different snapshots. The gray scale represents the projected H$_2$ density for the GMC. The contour colours represent the cores based on the global feedback bins (Sect. \ref{sec: global feedback bins}). We also visualise the changes in the properties of the cores by plotting them as spheres with their radius as the various properties (Fig. \ref{fig: core feedback prop}). By analysing the cores from multiple snapshots, we also follow the evolution of the simulated GMC and understand the effects of the evolutionary stage of the GMC on the core properties. Figure \ref{fig: core vis plots} shows the spherical GMC (top left) that appears to contract under self-gravity (top right) to form dense structures that lead to the formation of protostars. The GMC further evolves to have main-sequence stars (Fig. \ref{fig: core vis plots}, bottom left) with stellar winds, which along with other feedback mechanisms leads to a more scattered configuration of the cores. Figure \ref{fig: core vis plots} bottom right plot shows the GMC in which star formation has halted and supernova explosions have started. Here, most of the molecular gas is dispersed by the various feedback mechanisms.

Most of the cores appear to be elongated filamentary structures (Fig. \ref{fig: core vis plots}). These are projected images of the GMC and 3D plots might reveal different shapes for cores. However, it is beyond the scope of this paper.
Figure \ref{fig: core feedback prop} shows that most of the cores at later times belong to moderate and high feedback bins and have high velocity dispersions and virial parameters. 
This agrees with the results of \cite{xu2020ApJ...905..172X} that the outflows driven by older sources have a higher velocity dispersion. Moreover, stellar winds and supernova feedback, which start at later snapshots, also contribute to these velocity dispersion values.
The moderate and high-feedback bins also contain a few massive cores. These are bound cores with low-velocity dispersions that are likely sites for star formation.
The cores in the later snapshots appear more dispersed and separated. This is a result of the evolution and expansion of the GMC as seen in Fig. \ref{fig: core vis plots}.
    
\subsection{Core properties versus mechanical feedback}\label{subsec : results scatterplot}

\begin{figure*}
    \centering
    \includegraphics[width = \textwidth, keepaspectratio]{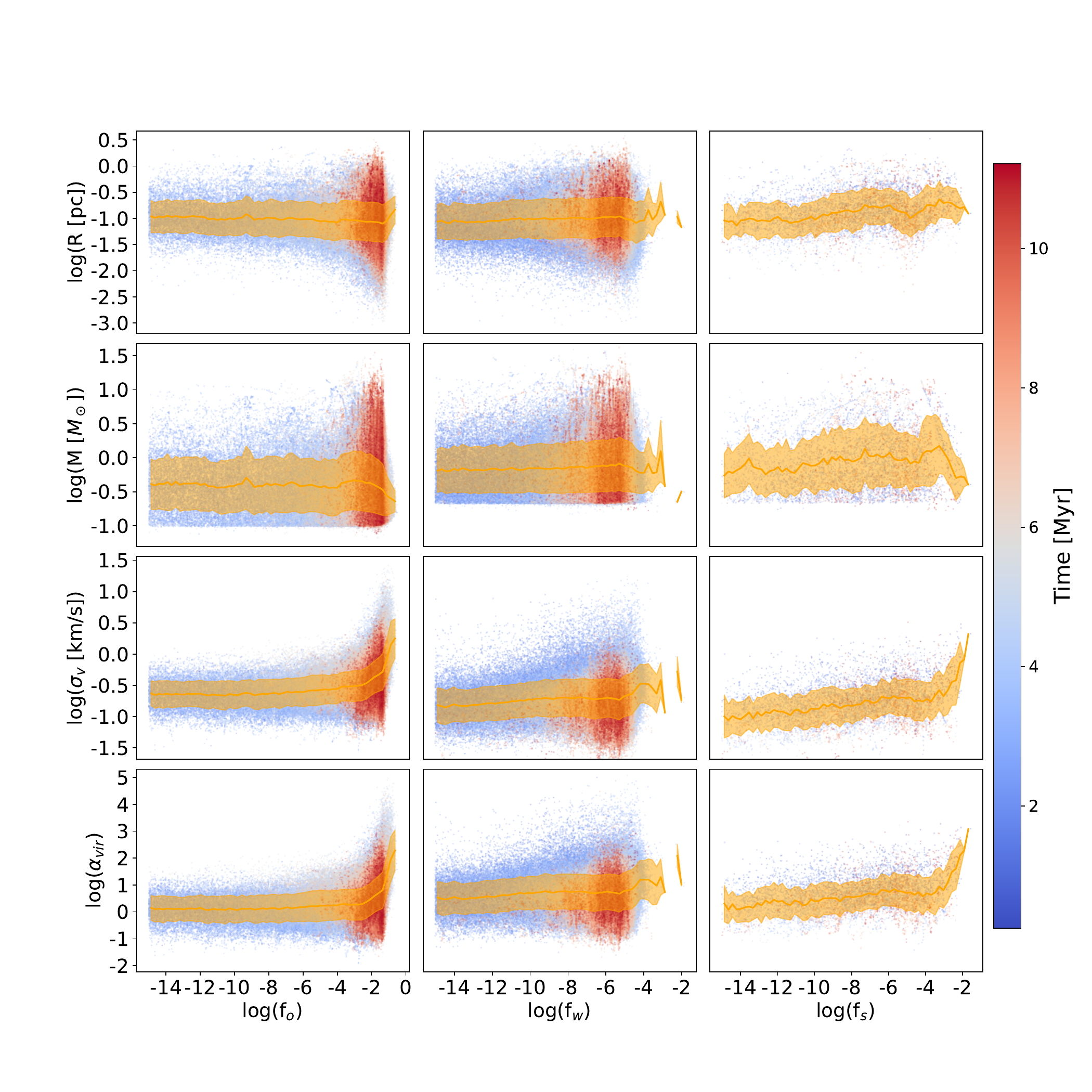}
    \caption{Core properties versus feedback fraction from outflows (left), winds (centre) and supernovae (right). The rows represent the radius of the core, the velocity dispersion, the mass, and the virial parameter, respectively. The orange filled area represents the average distribution of all cores, obtained by binning over the feedback fractions to get the 25th and 75th percentiles. The point colour indicates { the simulation time when the core is identified}, as time elapsed in Myr.}
    \label{fig: m2e4a2 scatter}
\end{figure*}

In this section, we compare the properties of the cores with the feedback fractions of protostellar outflow, stellar wind, and supernova. 
The distribution of core properties with feedback fractions <  $10^{-15}$ ( $\sim 10\%$) is similar to the property distribution with feedback fractions around $10^{-15}$.
While the feedback fraction is captured down to machine precision, there is little change in the core properties, and the regions/cores that have very low feedback fraction are not noticably affected by the feedback. The turbulent eddy mixing model implemented in {\small GIZMO} rapidly distributes the feedback gas throughout the domain as soon as feedback processes are initiated. Therefore, we truncate Figure \ref{fig: m2e4a2 scatter} to $10^{-15}$ to focus on the trends associated with the higher feedback fractions. 
The cores in the very late snapshots do not have the highest feedback fractions. This suggests that the feedback is not entirely cumulative. These later timescales represent the epoch when star formation has nearly halted, so the rate of feedback injection from new stars is significantly reduced \citep[star formation history of the simulated GMC][as shown in their Figure 2]{guszejnov2022}.

Figure \ref{fig: m2e4a2 scatter} shows an increase in the extreme values of the core radius\footnote{Core radii are calculated with an assumption that the cores are spherical, however, most of our cores appear to be filamentary. Thus, an elongated core also leads to a large effective core radius.} with the feedback fraction for both protostellar outflows and stellar winds. We find two competing effects.
The cores with the largest radii belong to the late-time snapshots, where a supernova explosion has already occurred. The increase in the core radii is most likely due to the momentum injected by the various feedback mechanisms. Due to the injected momentum, there could be expansion, rotation and shear motions in the cores, which increase their size. Figure \ref{fig: core vis plots} shows most cores as elongated or irregular structures rather than spheres. 
At the same time, feedback also compresses the cores and disperses the molecular gas, producing smaller cores. The smaller high-feedback cores could also be due to the evolving conditions in the cloud as the GMC collapses, which produces smaller cores while simultaneously increasing the number of stars and the amount of stellar feedback, which leads to higher feedback fractions in the cores. Despite the presence of large cores with high feedback, our overall analysis shows that outflows and winds lead to a decrease in the average core radius.

The cores with the largest velocity dispersions are in regions with significant stellar feedback. Figure \ref{fig: m2e4a2 scatter} shows a gradual increase in the average velocity dispersion of the cores as the feedback increases, and a rapid increase in the cores with high feedback fractions. The increase in velocity dispersion for cores with high feedback is likely due to the momentum injected by the feedback mechanisms \citep{offner2017}. 
The most turbulent and least turbulent cores have high wind feedback fractions ($f_{\rm{w}}$) (Fig. \ref{fig: m2e4a2 scatter}; central plot in the second row). The cores with small velocity dispersions belong to the evolved GMC where star formation has halted and the amount of gas from stellar winds is lower compared to the peak star formation phase.

The third row of panels in Fig. \ref{fig: m2e4a2 scatter} shows that the average mass distribution is generally constant and only decreases at high feedback fractions ($f_{\rm{o}}$, $f_{\rm{w}}$, and $f_{\rm{s}}$).
The decrease in mass at high feedback fractions might be the result of feedback destroying cores and splitting them into less massive cores. 
However, the most massive cores have relatively high feedback fractions and belong to late-time snapshots. These cores are most likely the result of winds and supernova compressing the gas.

The bottom row of Figure \ref{fig: m2e4a2 scatter} shows a gradual increase in the average virial parameter with increasing feedback fraction; with a strong increase at the highest feedback fractions. It is clear that this is driven by higher velocity dispersions, as the core mass and radius do not change significantly.

\subsection{Violinplots}\label{sec : violinplots}

\begin{figure*}
    \centering
    \includegraphics[width = 0.49\textwidth, keepaspectratio]{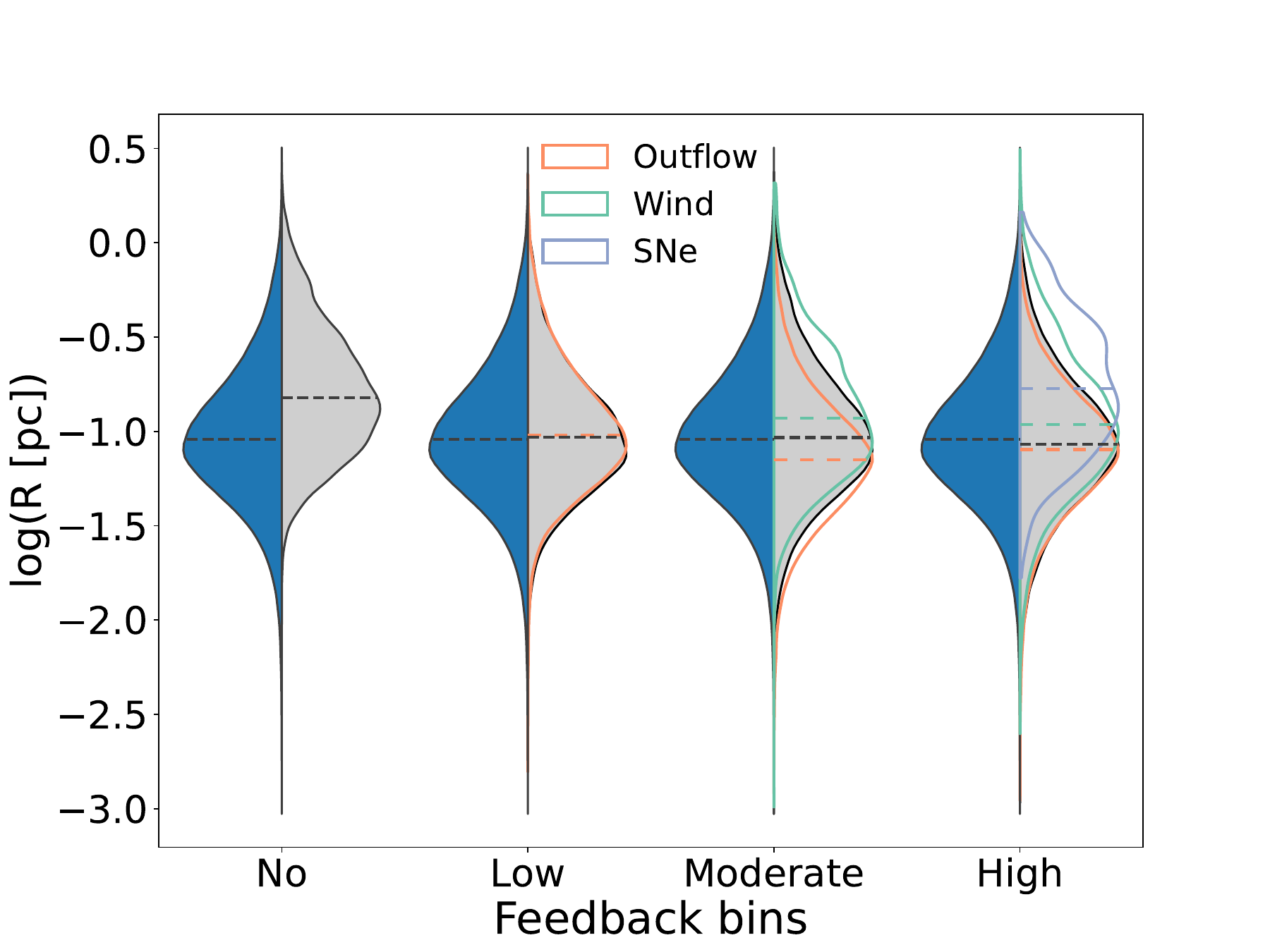}
    \includegraphics[width = 0.49\textwidth, keepaspectratio]{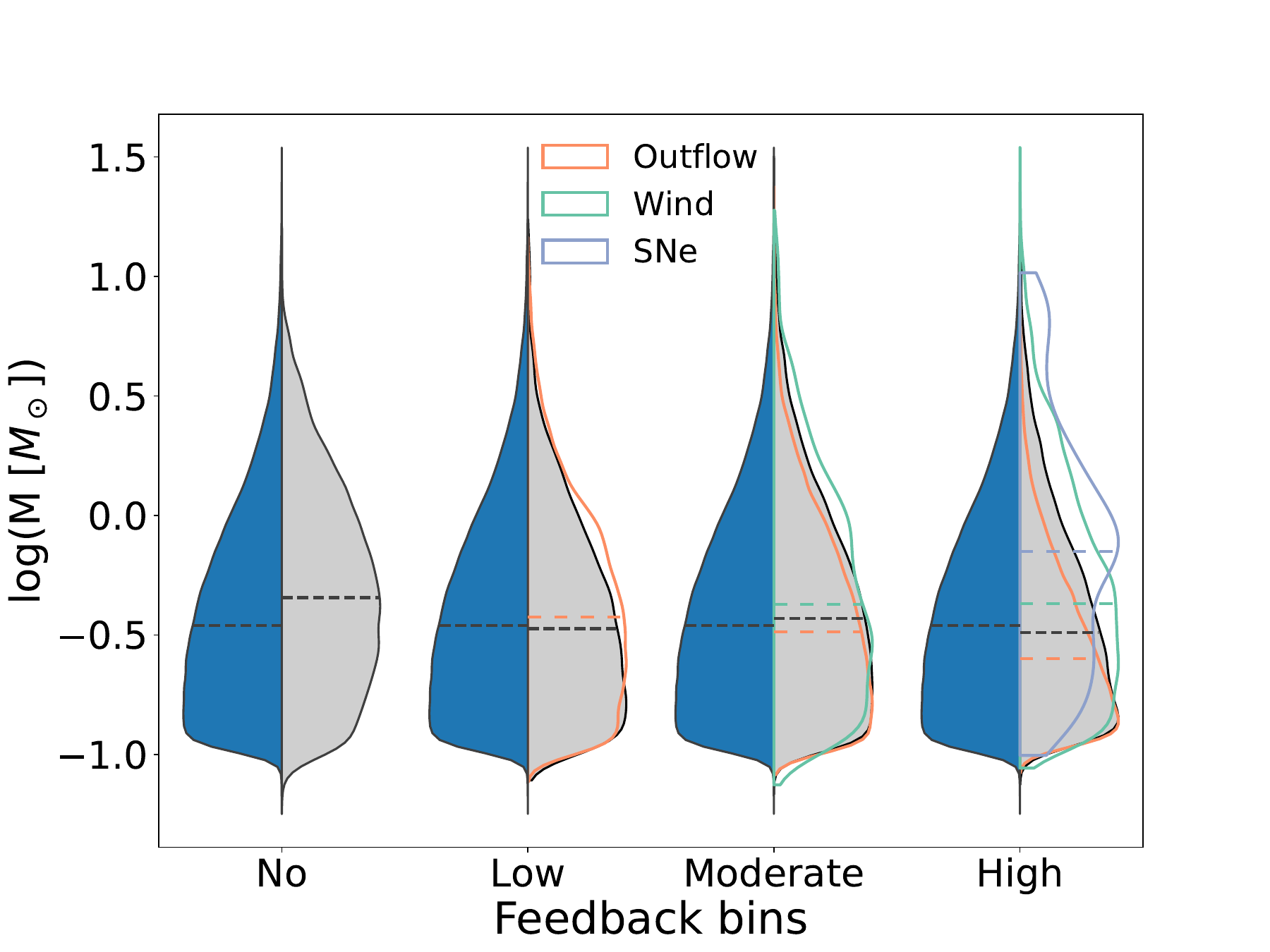}
    \includegraphics[width = 0.49\textwidth, keepaspectratio]{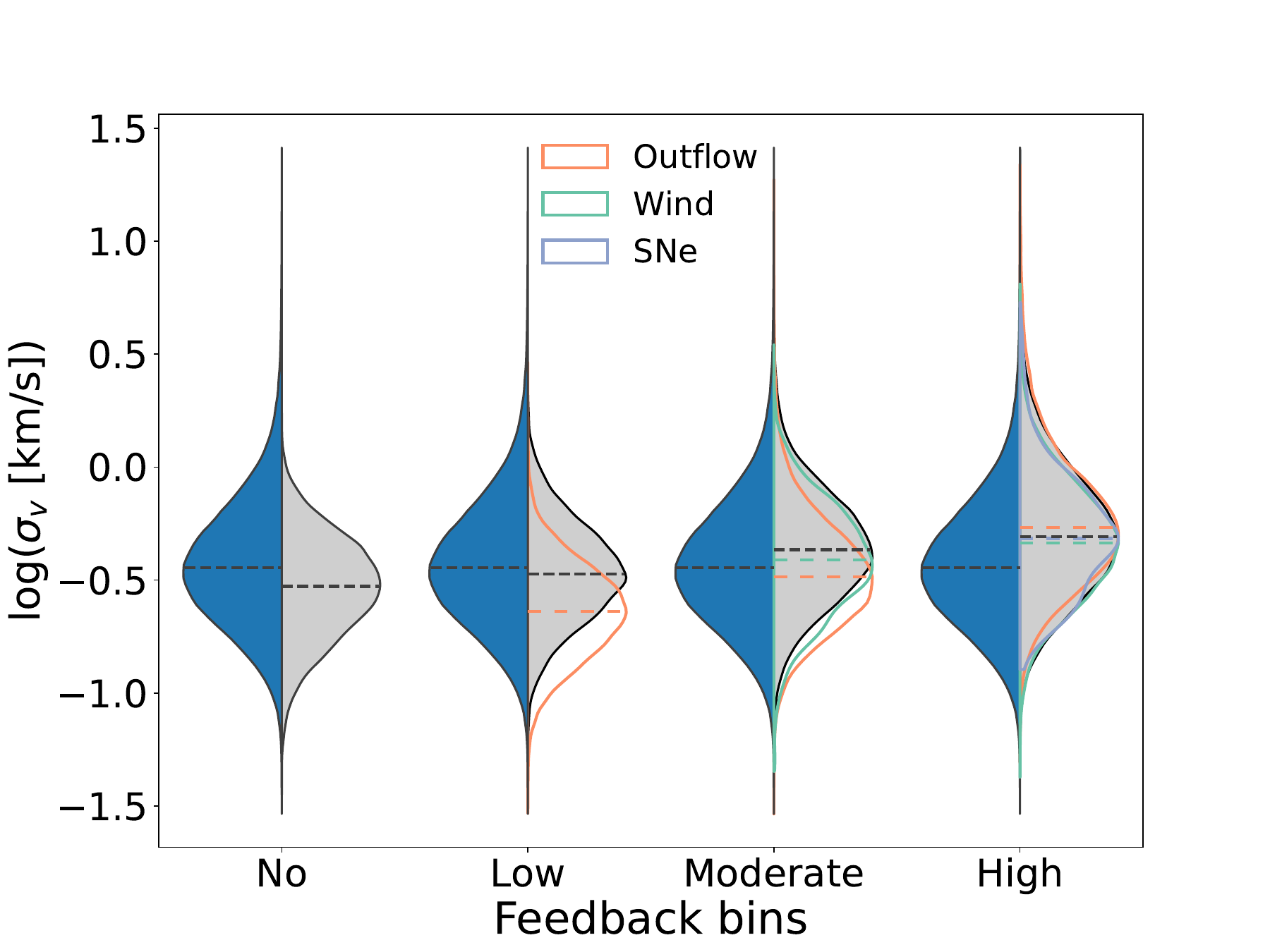}
    \includegraphics[width = 0.49\textwidth, keepaspectratio]{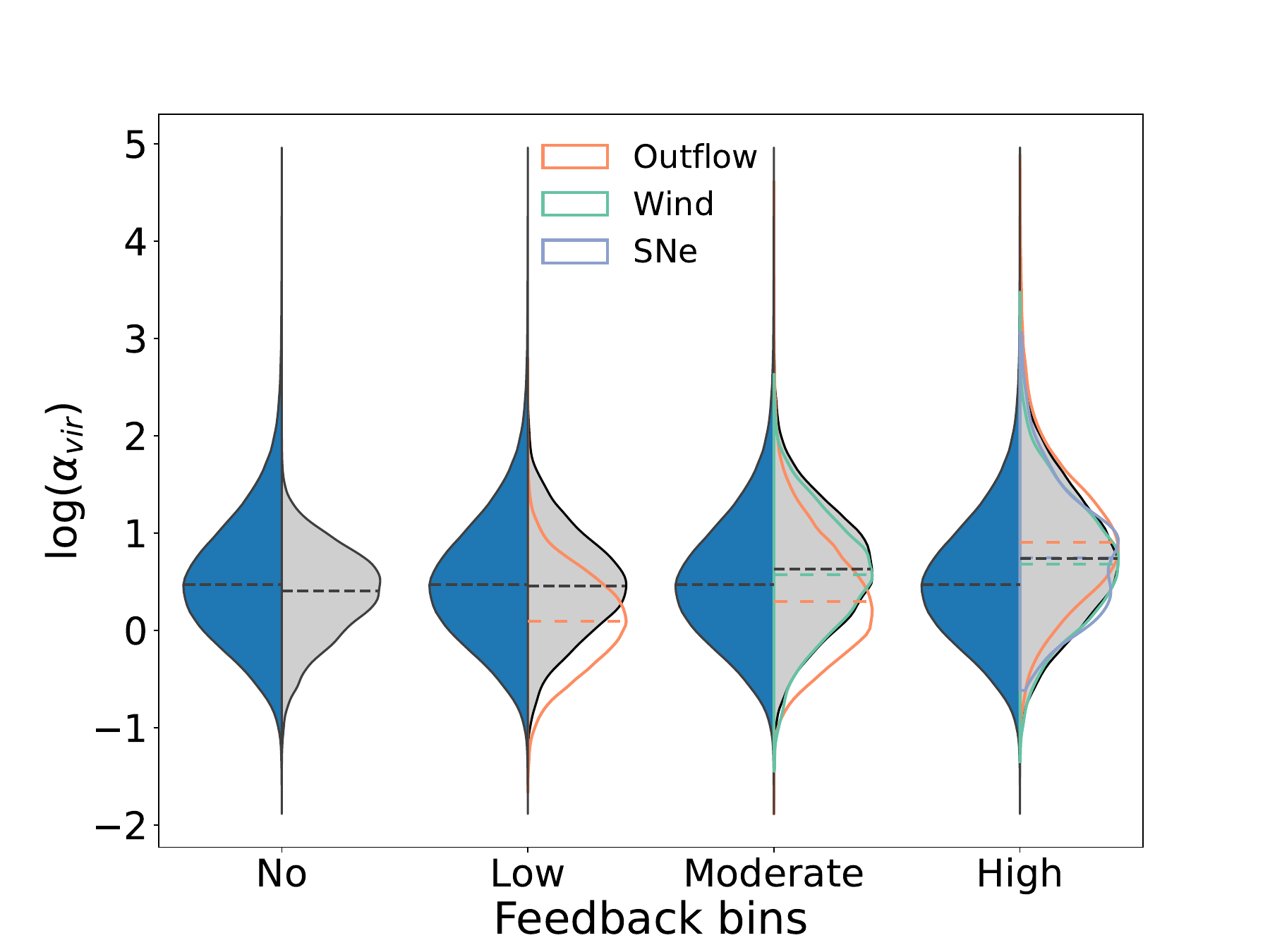}
    \caption{Distribution of core radius (top left), mass (top right), velocity dispersion (bottom left) and virial parameter (bottom right) in the different feedback bins. The split violin plot represents the total distribution (blue) on the left and the distribution of cores in the global feedback bins (grey) on the right for each axis. The colours representing outflow, wind, and supernova bins are marked in the legends. The dashed lines represent the medians of the different distributions. Violin plots present the density of the data at different values, which is smoothed through the kernel density estimator.}
    \label{fig: m2e4a2 feedback violin}
\end{figure*}

In this section, we use double violin plots (Fig. \ref{fig: m2e4a2 feedback violin}) to visualise the distributions of core properties in the various feedback bins defined in Sect. \ref{sec: three feedback bins} and \ref{sec: global feedback bins}. Binning the data using percentiles based on the feedback fraction allows us to study the changes in the properties for the cores that are actually influenced by stellar feedback. 
The low wind, low supernova, and moderate supernova bins contain almost no cores (Table \ref{tab: cores in bins}) and are therefore not included in the violinplots.
We use the Wasserstein distance (App. \ref{sec: wasserstein distance}) to statistically compare the different distributions. 

Figure \ref{fig: m2e4a2 feedback violin} shows a decrease in the average radius and mass from the no-feedback bin to the low global feedback bin. 
However, the radius 
(Fig. \ref{fig: m2e4a2 scatter}) shows only a slight change from low to high global feedback. The mass distributions show negligible changes over the feedback bins.
The cores in the low global bin are most likely the star-forming cores that are gravitationally collapsing and therefore are smaller (Fig. \ref{fig: core vis plots}).
This suggests that the main impact of the feedback occurs via turbulence driving / energy injection. The lower average values of 
radius in the outflow bins compared to the respective wind and supernova bins point to the disruptive effects of these mechanisms on the cores \citep{fuente2002A&A...387..977F, maret2009ApJ...698.1244M, narayanan2012MNRAS.425.2641N}. However, flows that play a very important role in core dispersal may not be the main factor in GMC disruption \citep{maret2009ApJ...698.1244M, guszejnov2022}.
The nearly constant radius and mass over the wind bins suggest that stellar winds do not affect the radius of the cores as strongly as other mechanisms or that the expansion of the cores due to stellar winds might be suppressed due to the dissipation of gas by the various feedback mechanisms \citep[including winds themselves, e.g.,][]{pabst2019Natur.565..618P}. The high supernova bin shows an average mass and radius higher than any other bins, including the pristine cores (grey violins in Fig. \ref{fig: m2e4a2 feedback violin}) and the total core distribution (blue violins in Fig. \ref{fig: m2e4a2 feedback violin}). These cores are likely due to gas compression along the shock front of the supernova explosion, leading to elongated massive structures.

The lower panels of Figure \ref{fig: m2e4a2 feedback violin} show the velocity dispersion and virial parameter distributions, which both montonically increase from no- to high-feedback bins. 
The increase in core velocity dispersion from low- to high-feedback bins for all feedback mechanisms (Fig. \ref{fig: m2e4a2 feedback violin}) is consistent with the standard picture of momentum/energy injection via stellar feedback \citep{arce2007prpl.conf..245A, offner2017, offner2018, bieri2023MNRAS.523.6336B, geen2023}. Observations of MCs have also revealed an increase in the line widths of structures in feedback-dominated regions \citep{wong2022ApJ...932...47W, grishunin2024A&A...682A.137G}.
The low-outflow bin velocity dispersion distribution is significantly different from the distribution in the low-global feedback bin. This difference becomes smaller in the moderate and high feedback bins.
This large increase in velocity dispersion 
is likely a result of outflows injecting significant momentum and energy, 
thereby becoming a major source of turbulence in the cores \citep{zhang2005ApJ...625..864Z, arce2007prpl.conf..245A, offner2017,davis2008MNRAS.387..954D, duarte-cabral2012A&A...543A.140D, bally2016, verliat2022A&A...663A...6V}.
In comparison to outflows, stellar winds appear to have less influence on velocity dispersion, which increases less in cores dominated by wind feedback (Sect. \ref{sec : violinplots}).
Although the total integrated wind energy injection may be higher in GMCs like the one we model \citep{grudic2022}, the wind energy and momentum are injected at later times by massive stars into lower-density gas, occurring after much of the dense core envelope has already been accreted or dispersed.
Previous studies also show that winds deposit energy and momentum locally \citep{fichtner2022MNRAS.512.4573F} but do not offset global turbulence in a molecular cloud \citep{offner2015ApJ...811..146O, guszejnov2022}.

Figure \ref{fig: m2e4a2 feedback violin} (bottom right) shows that the core virial parameters span three orders of magnitude.
Although the typical core 
is close to virial equilibrium, many cores have significantly higher virial parameters. The large spread in Figure \ref{fig: m2e4a2 feedback violin} (bottom right) is not unexpected, since we do not require boundedness for core identification and many of the identified cores likely do not go on to form stars \citep[e.g.,][]{offner2022MNRAS.517..885O}. We expect that a larger fraction of cores will appear bound for choices of higher {\it min\_value}. However, these results are not inconsistent with what we find for observed cores and clumps, which exhibit a broad range of virial parameters ranging from 0.03 to a few hundred \citep{kauffmann2013ApJ...779..185K}.
Here, the cores with high virial parameters could be pressure-dominated structures such as low-density thermal cores \citep{lada2008ApJ...672..410L}, droplets \citep{chen2019}, or overdensities in shocked regions.

\subsection{Scaling relations}\label{subsec: scaling relations}

In this section, we analyse the correlations between the core properties for various feedback bins and compare them to three scaling relations from the star formation literature. The core distributions in scaling relation plots are shown with 1-sigma kernel density estimators (KDEs)\footnote{We use the seaborn KDE plots \url{https://seaborn.pydata.org/generated/seaborn.kdeplot.html} with level = 0.0027, which corresponds to 1-sigma contours.}.
We also use principal component analysis (PCA), following \cite{colombo2019} and \cite{ neralwar2022b}, to obtain the slopes of the distributions and their scatter.

PCA constructs the covariance matrix between the two core properties and calculates the eigenvectors of the matrix. The inclination of the largest eigenvector (major axis of the PCA ellipse) with respect to the x-axis gives the slope of the distribution. The eigenvector perpendicular to the major axis is a measure of the scatter in the data. 
The errors on slopes and scatter (Tables \ref{tab: larson slopes and width} - \ref{tab: size mass slopes and width}) were obtained using the bootstrap technique, where 1000 new samples are generated from the main sample by drawing with replacement. The slopes and widths in the table represent the respective mean values of the bootstraped samples. $\sigma_{\rm{slope}}$ and $\sigma_{\rm{width}}$ provide the errors on the slopes and widths, respectively; calculated as the standard deviation of the respective values on the bootstraped samples. 
While scaling relations have been measured extensively in observations of MCs, here we study cores in simulations. Moreover, we obtain the properties of the cores using methods different from those that observers generally use. Observed structures, particularly those defined in optically thick molecular line tracers like $^{12}$CO and $^{13}$CO,  are subject to projection and opacity effects that may significantly impact inferred masses, velocity dispersions and virial parameters \citep{beaumont2013ApJ...777..173B, mairs2014ApJ...783...60M}. We caution the reader that this may lead to discrepancies in the slopes for the different distributions. 

Larson's size-linewidth relation $\sigma_\varv$ = 1.10 $L^{0.38}$ has been used to study and constrain the fundamental properties of MCs for the past 40 years \citep{larson1981}. It was obtained by studying MCs of sizes ranging from sub-pc scales to $\sim$ 100 pc size. Observations of dense cores \citep{myers1983}, and MCs \citep{solomon1987} suggest a slope of 0.5, which is consistent with the latest observations \citep[e.g.][]{colombo2019, neralwar2022b}.

\begin{figure*}
    \centering
    \includegraphics[width = .47 \textwidth, keepaspectratio]{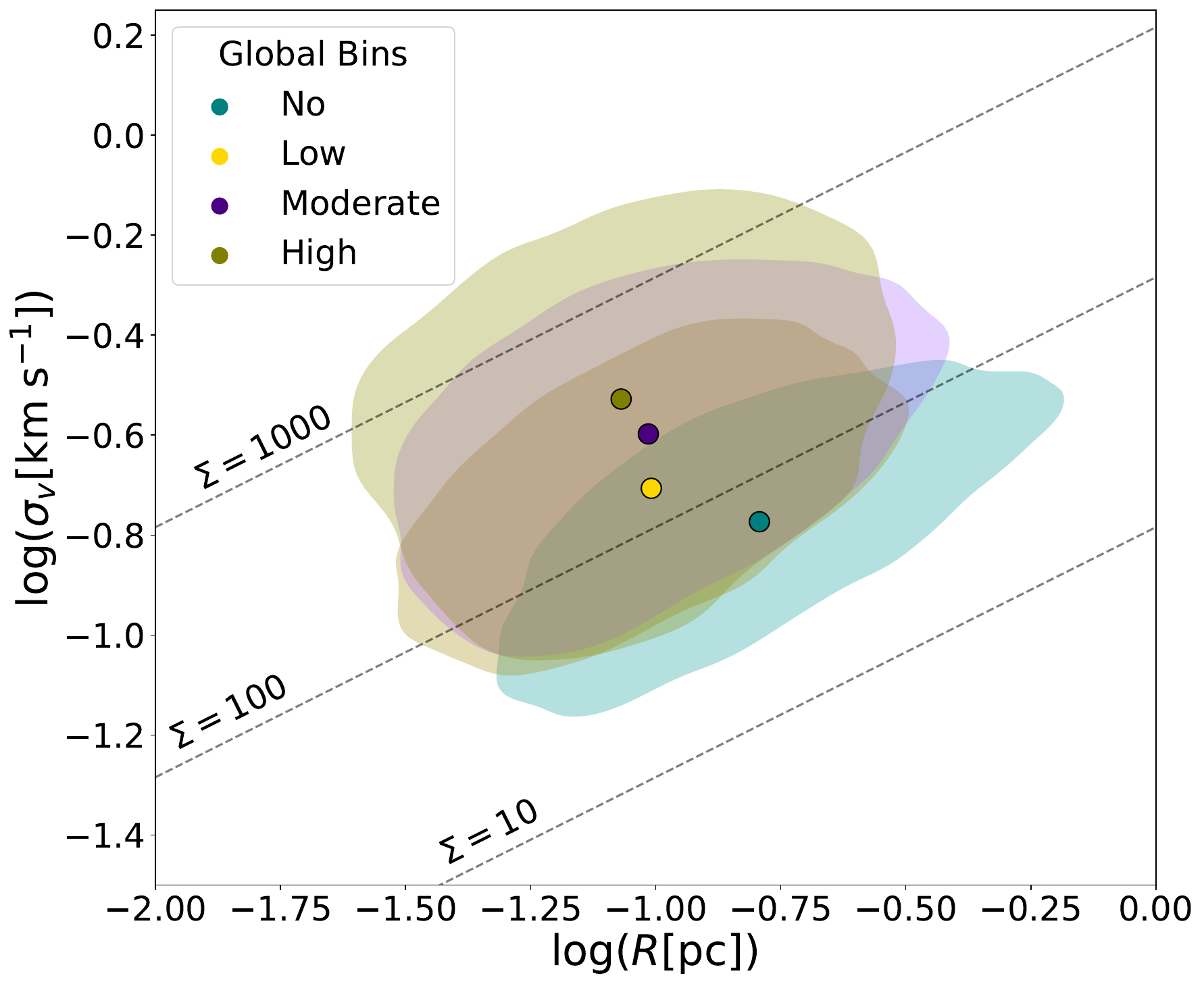}
    \includegraphics[width = .47 \textwidth, keepaspectratio]{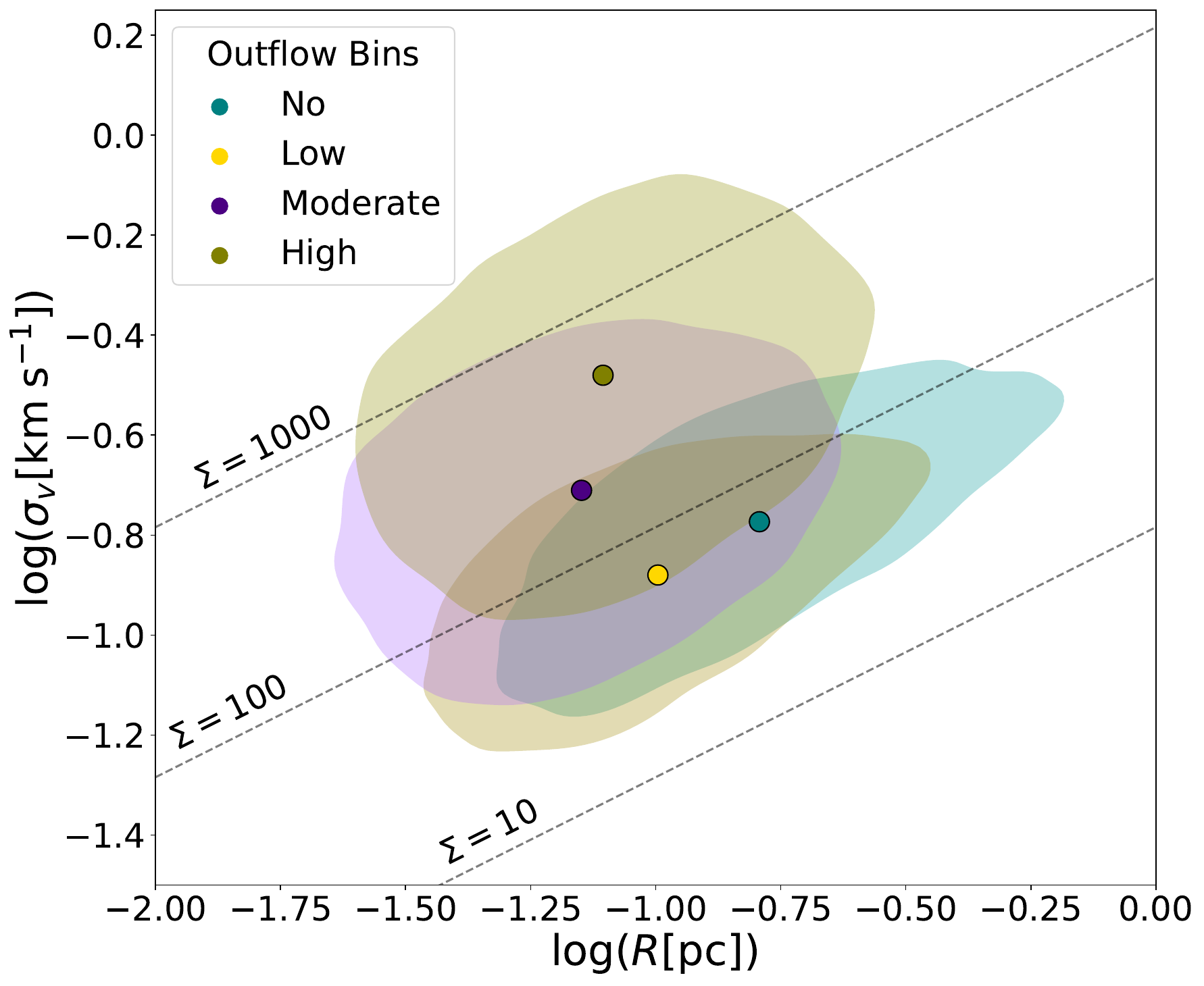}
    \includegraphics[width = .47 \textwidth, keepaspectratio]{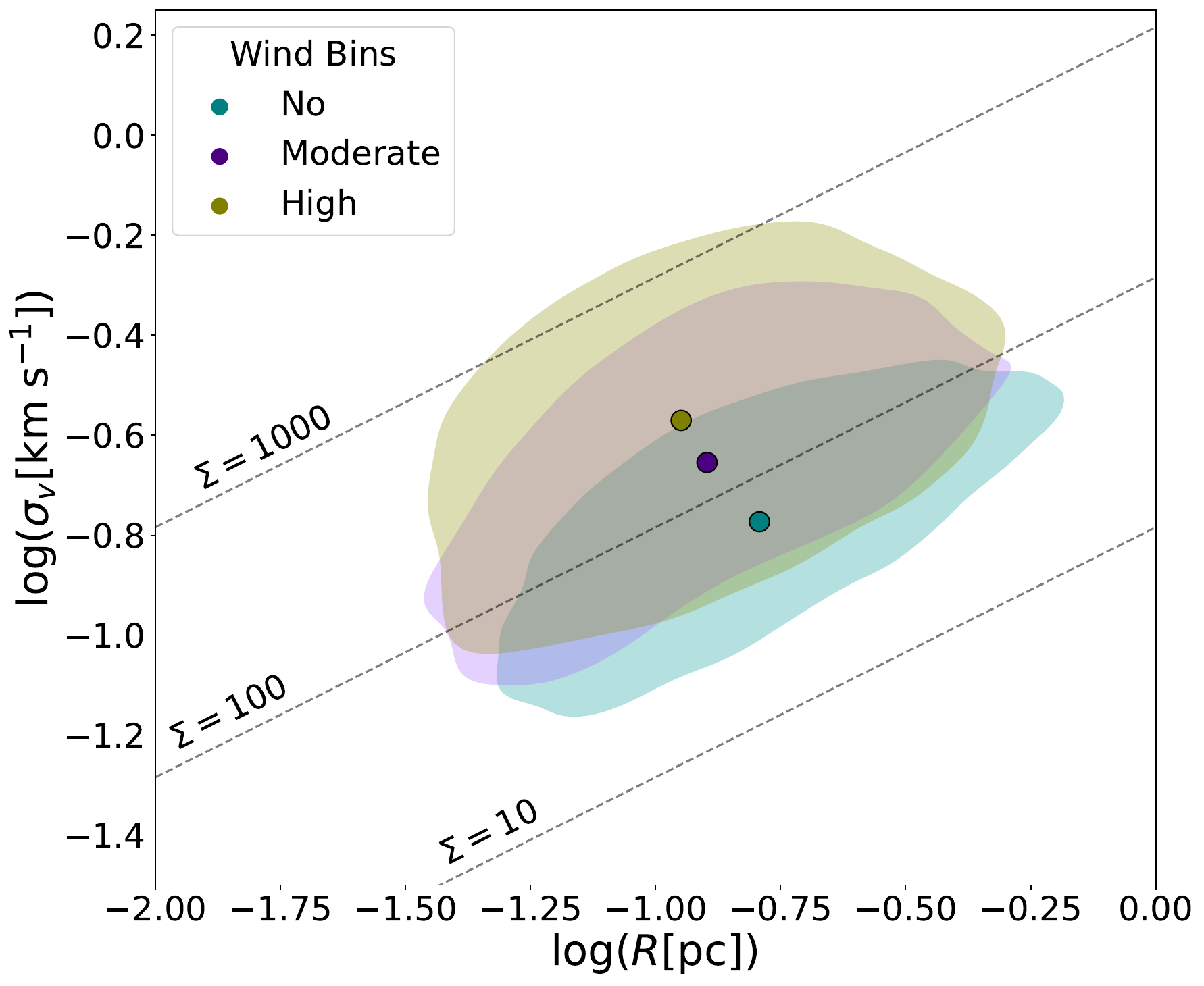}
    \includegraphics[width = .47 \textwidth, keepaspectratio]{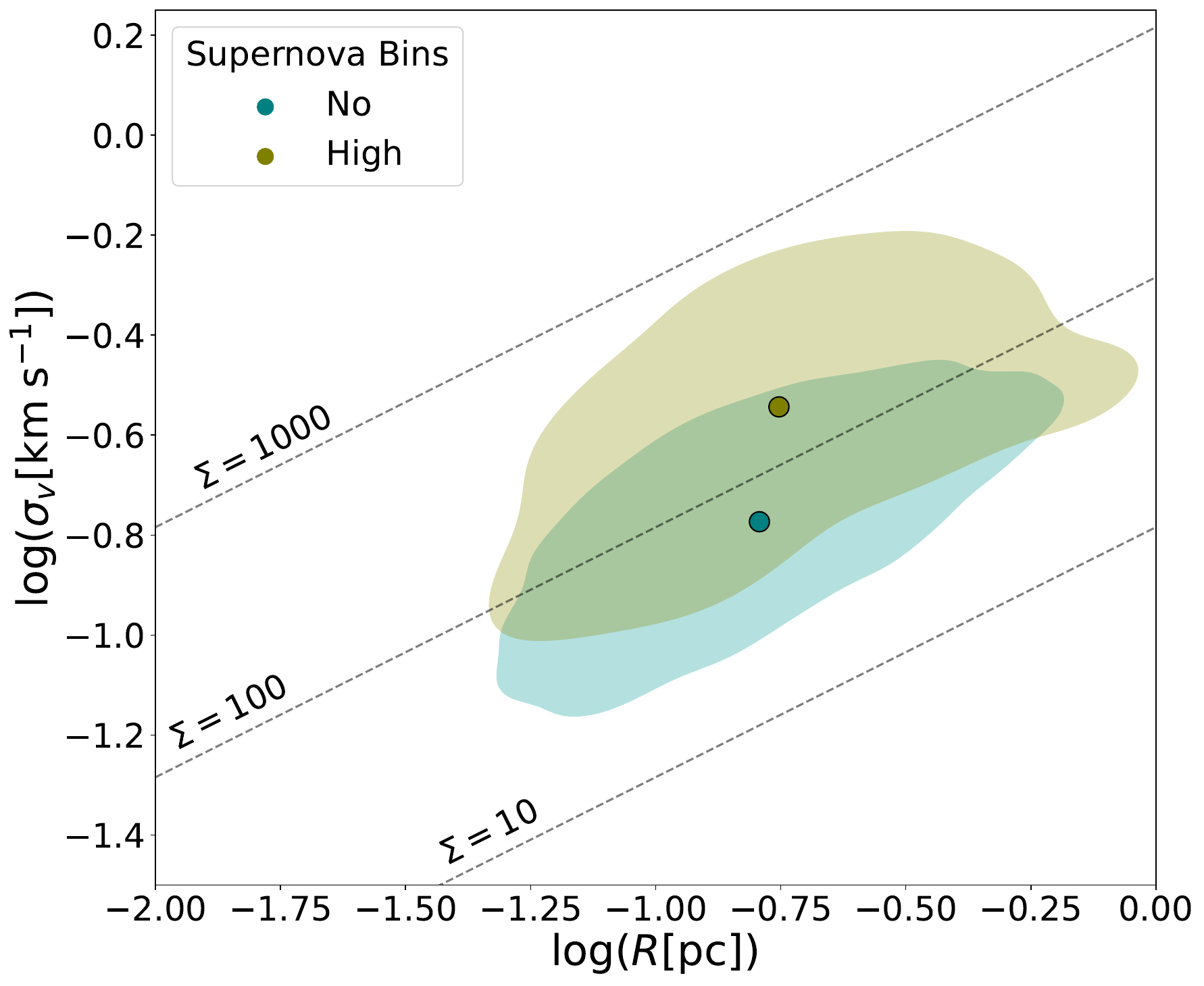}
    \caption{Size-linewidth relation ($\sigma_\varv$ versus $R$) for cores in different feedback bins. The top left plot represents the distibution in the four global feedback bins. The top right plot represents the distributions in the outflow bins. The lower left and lower right plots represent the wind and supernova bins, respectively.
    The dashed lines represent Larson's first relation \citep[][, slope of 0.5]{larson1981, solomen1969} for different surface mass densities (units of $\rm{M}_\odot \rm{pc}^{-2}$) for virialised structures ($\alpha_{\rm{vir}} = 1$). The different colours represent the distribution of cores in the various feedback bins, as noted in the legend.}
    \label{fig: m2e4a2 larson feedback}
\end{figure*}

Figure \ref{fig: m2e4a2 larson feedback} presents Larson's law for cores in different feedback bins. A general observation is that the cores in each bin follow Kolmogorov's turbulence law, i.e., larger cores have a higher velocity dispersion. All sets of cores follow the expected observational trend; however, the cores in the higher feedback bins have a higher velocity dispersion compared to the similar-sized cores in the lower feedback bins. In addition, the cores in the no-feedback bin show a tighter correlation (Table \ref{tab: larson slopes and width}).

Figure \ref{fig: m2e4a2 larson feedback} (top right)  shows a shift in the distribution from no- to low-outflow bins towards smaller cores with low-velocity dispersions. These low-outflow cores are detected in the early stages of GMC evolution, when denser structures are still forming, and parts of the GMC are collapsing under self-gravity. The moderate- and high-outflow bins represent a distribution of cores with strong outflows leading to gas dispersal and high-velocity dispersion.
The distribution of high supernova feedback cores (Fig. \ref{fig: m2e4a2 larson feedback}; bottom right) contains some of the largest cores in the simulation, which is likely due to the momentum from the supernova leading to core expansion and/or elongation. 
Many of these cores are identified in the large, high-velocity shell of entrained gas created by the supernova shock.
However, the cores with the highest velocity dispersion are present in the high-outflow and high-wind bins rather than the high-supernova bin. This suggests that a supernova may be less efficient in driving the turbulence within the dense gas. This agrees with the notion that outflows and winds deposit energy more locally, directly into the dense gas where the stars are forming.

\begin{table}[]
\caption{Larson relation: slopes and widths of PCA ellipse.}
\label{tab: larson slopes and width}
\centering
\begin{tabular}{lllll}
\hline
Feedback Bins & Slope & $\sigma_{\rm{slope}}$ & Scatter & $\sigma_{\rm{scatter}}$ \\ \hline
No            & 0.54  & < 0.01                  & 0.3     & < 0.01                    \\ \hline
\multicolumn{5}{c}{Global}       \\
Low           & 0.48  & 0.01                  & 0.34    & < 0.01                    \\
Moderate      & 0.38  & < 0.01                  & 0.46    & < 0.01                    \\
High          & 0.35  & 0.01                  & 0.58    & < 0.01                    \\ \hline
\multicolumn{5}{c}{Outflow}       \\
Low           & 0.45  & < 0.01                  & 0.32    & < 0.01                    \\
Moderate      & 0.31  & 0.01                  & 0.48    & < 0.01                    \\
High          & 0.56  & 0.02                  & 0.56    & < 0.01                    \\ \hline
\multicolumn{5}{c}{Wind}       \\
Moderate      & 0.53  & 0.01                  & 0.36    & < 0.01                    \\
High          & 0.36  & 0.01                  & 0.48    & < 0.01                    \\ \hline
\multicolumn{5}{c}{Supernova}       \\
High          & 0.55  & 0.04                  & 0.42    & 0.01                    \\ \hline
\end{tabular}
\tablefoot{All values are rounded to two decimal places. The lines of constant surface densities in Fig. \ref{fig: m2e4a2 larson feedback} represent a slope of 1.}
\end{table}

\begin{figure*}
    \centering
    \includegraphics[width = .47 \textwidth, keepaspectratio]{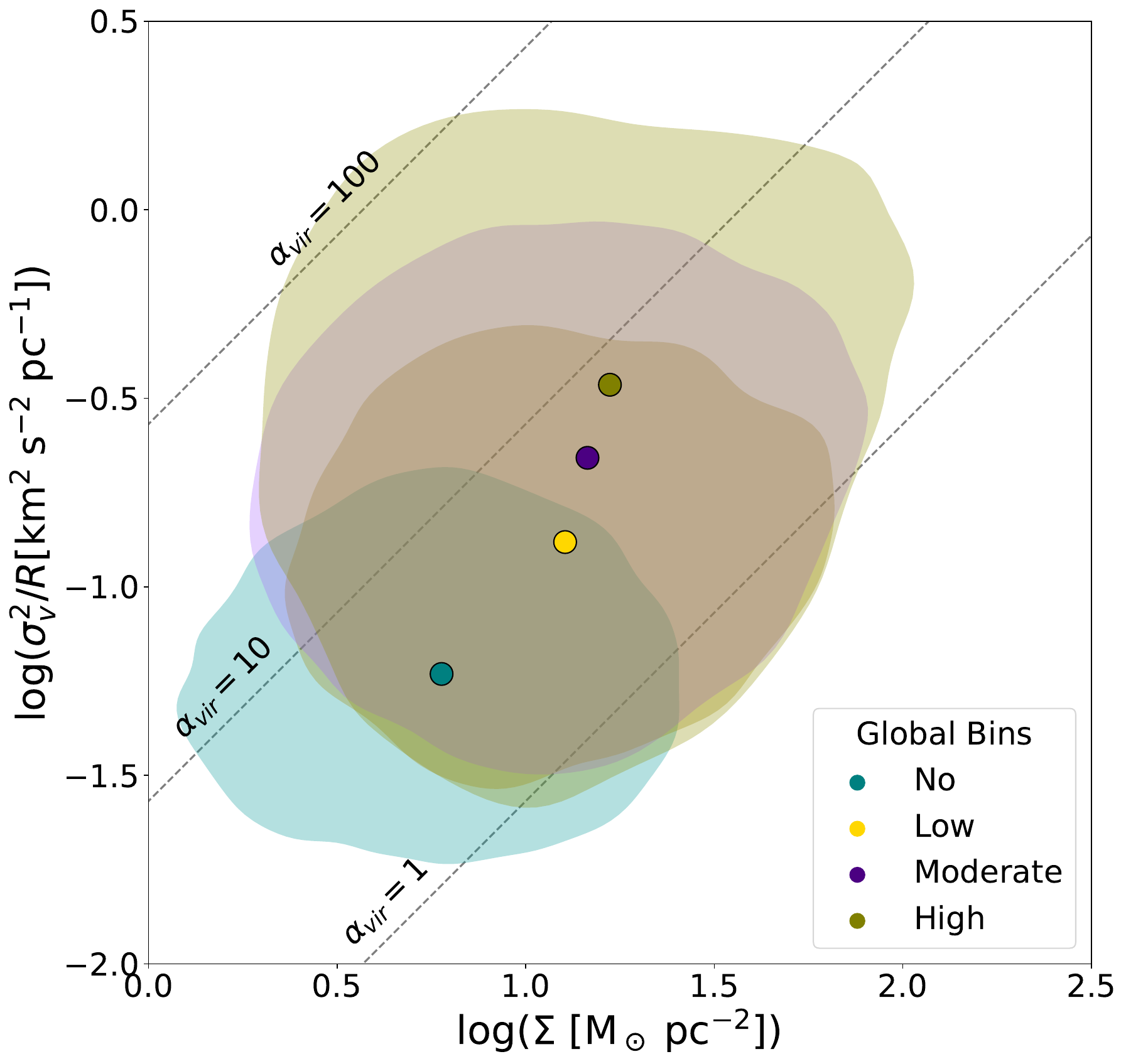}
    \includegraphics[width = .47 \textwidth, keepaspectratio]{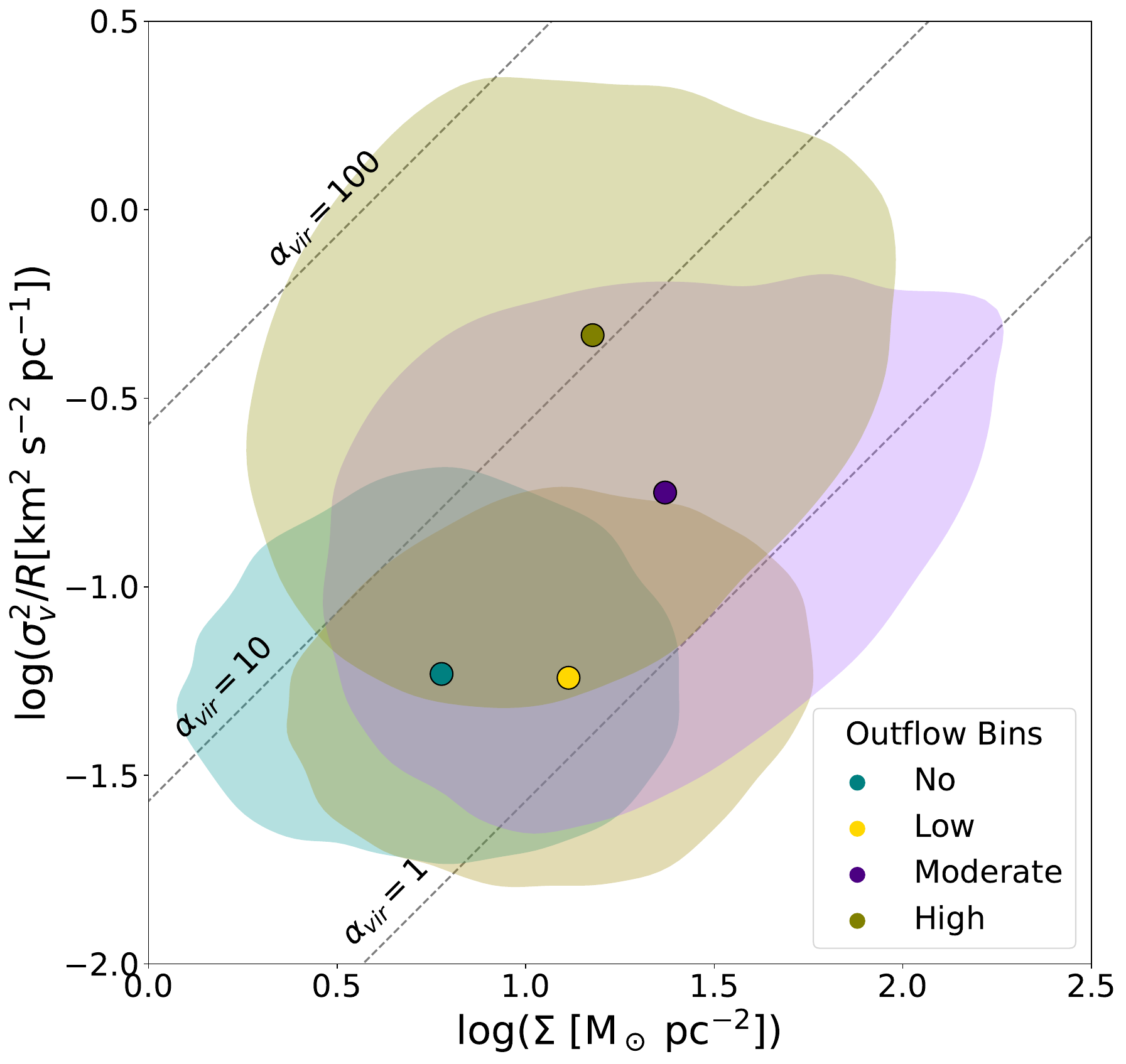}
    \includegraphics[width = .47 \textwidth, keepaspectratio]{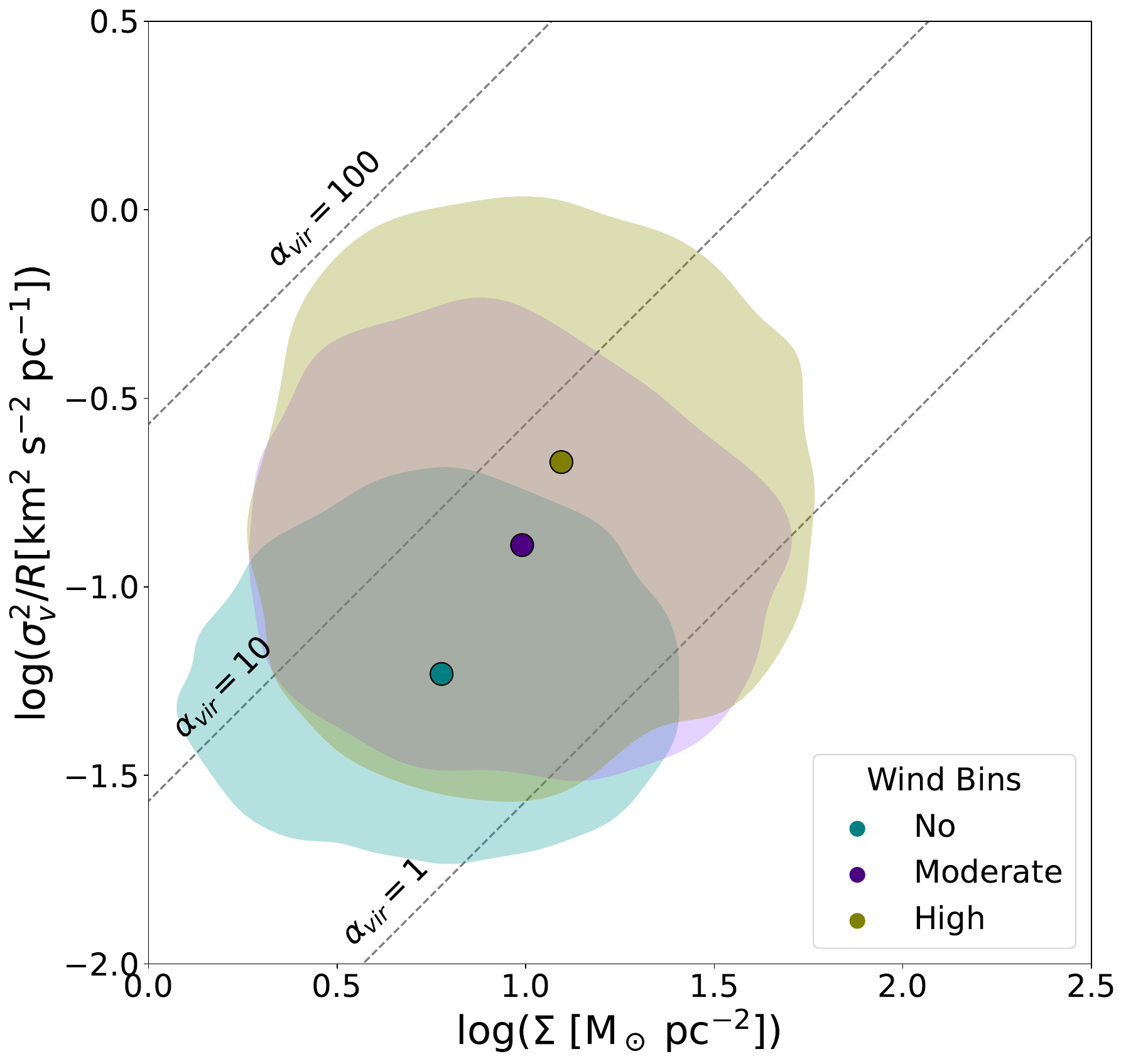}
    \includegraphics[width = .47 \textwidth, keepaspectratio]{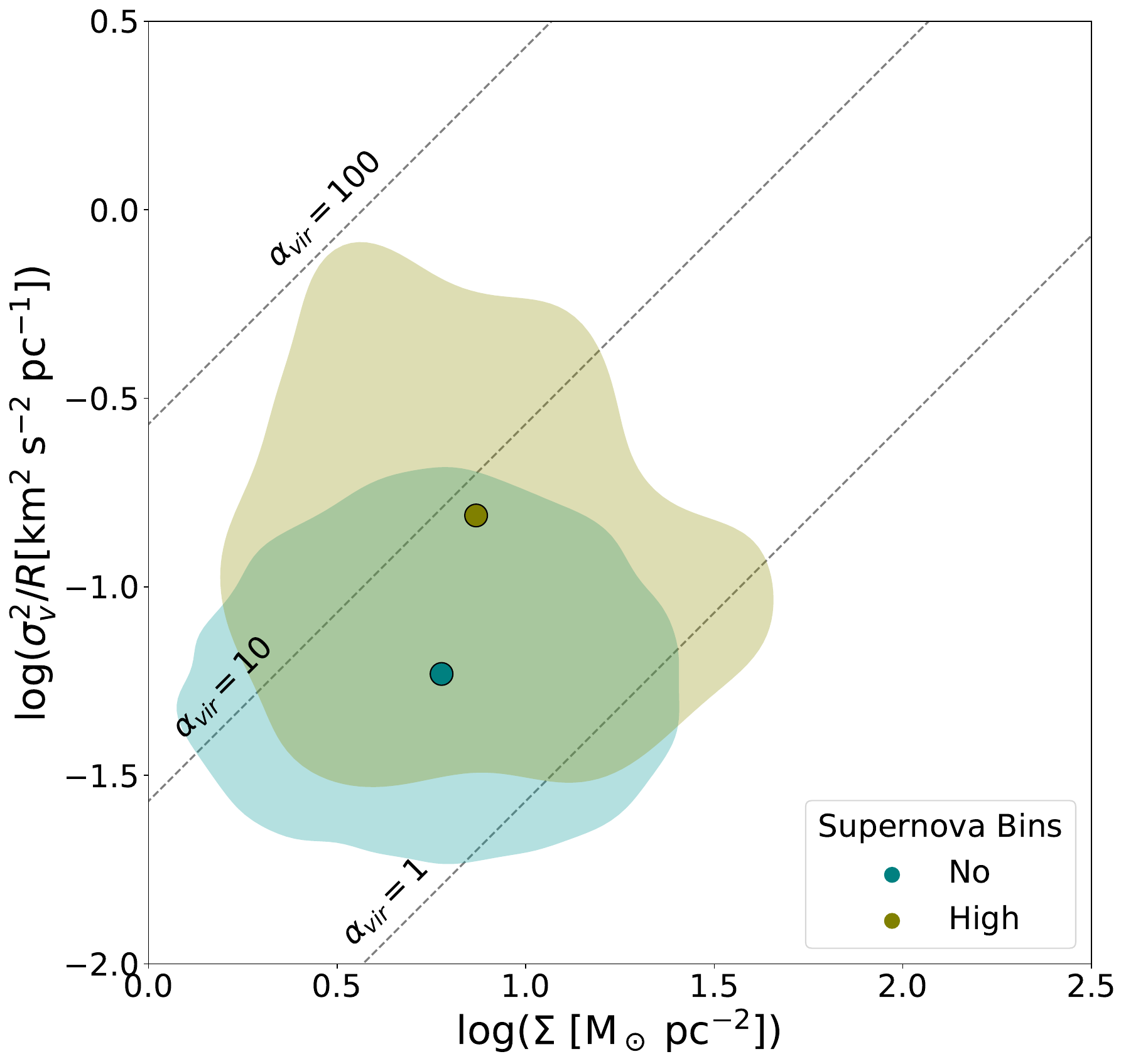}
    \caption{Scaling relation between $\sigma_\varv^2$/$R$ and gas surface density ($\Sigma$) for the cores in different outflow bins. The dashed lines are isocontours of virial parameters. The symbols and conventions for the subplots follow Fig. \ref{fig: m2e4a2 larson feedback}.}
    \label{fig: m2e4a2 heyer feedback}
\end{figure*}

The Heyer relation \citep{heyer2009} ($\sigma_\varv^2/R \propto \Sigma$) compares the surface density of structures with their scaling parameter (radius and velocity dispersion). Fig. \ref{fig: m2e4a2 heyer feedback} shows that the cores impacted by any type of feedback have higher typical surface densities compared to the pristine cores. 
This is consistent with significant feedback injection occurring while stars are still embedded. Supernova and stellar wind feedback may also compress gas, leading to higher densities.
The cores in low- and moderate-outflow bins have a higher surface mass density but comparable virial parameters to pristine cores.
The collapse of the GMC under self-gravity in early times with low feedback leads to these dense bound cores.
The structures in higher feedback bins have higher virial parameters, i.e., are more unbound. As shown in Sect. \ref{sec : violinplots},  this is largely because stellar feedback increases velocity dispersion in these cores.

Our cores have relatively low surface mass densities compared to dense cores found in observations \citep{ballesteros-paredes2020}. However, our surface densities are not obtained in a fashion similar to those obtained by observations. We calculated the mass and the area of the cores based on their actual 3D structure and not the ppv cubes used in observational studies that provide only the projected images of clouds. Since we identify cores using the 3d density field, our estimated cores are free from projection effects and line-of-sight contaminations \citep{beaumont2013ApJ...777..173B}.
Moreover, we analyse the total gas distribution, unlike most observational results on GMCs. Observations are often based on detection using molecular tracers (e.g. $^{12}$CO, $^{13}$CO) and dust emission, but do not probe the most diffuse molecular gas \citep{duarte-cabral2016MNRAS.458.3667D}.

\begin{table}[]
\caption{Heyer relation: slopes and widths of PCA ellipse.}
\label{tab: heyer slopes and width}
\centering
\begin{tabular}{lllll}
\hline
Feedback Bins & Slope & $\sigma_{\rm{slope}}$ & Scatter & $\sigma_{\rm{scatter}}$ \\ \hline
No            & 0.15  & 0.01                  & 0.67    & < 0.01                    \\ \hline
\multicolumn{5}{c}{Global}       \\
Low           & 0.31  & 0.01                  & 0.74    & < 0.01                    \\
Moderate      & 0.71  & 0.01                  & 0.89    & < 0.01                    \\
High          & 1.18  & 0.02                  & 1.04    & < 0.01                    \\ \hline
\multicolumn{5}{c}{Outflow}       \\
Low           & 0.4   & 0.01                  & 0.66    & < 0.01                    \\
Moderate      & 0.8   & 0.01                  & 0.87    & < 0.01                    \\
High          & 1.06  & 0.02                  & 1.03    & < 0.01                    \\ \hline
\multicolumn{5}{c}{Wind}       \\
Moderate      & 0.02  & 0.04                  & 0.81    & 0.01                    \\
High          & 0.89  & 0.02                  & 0.93    & 0.01                    \\ \hline
\multicolumn{5}{c}{Supernova}       \\
High          & -1.84 & 11.57                 & 0.89    & 0.03                    \\ \hline
\end{tabular}
\tablefoot{The lines of constant virial parameters in Fig. \ref{fig: m2e4a2 heyer feedback} represent a slope of 1. The symbols and conventions follow Table \ref{tab: larson slopes and width}.}
\end{table}

\begin{figure*}
    \centering
    \includegraphics[width = .47 \textwidth, keepaspectratio]{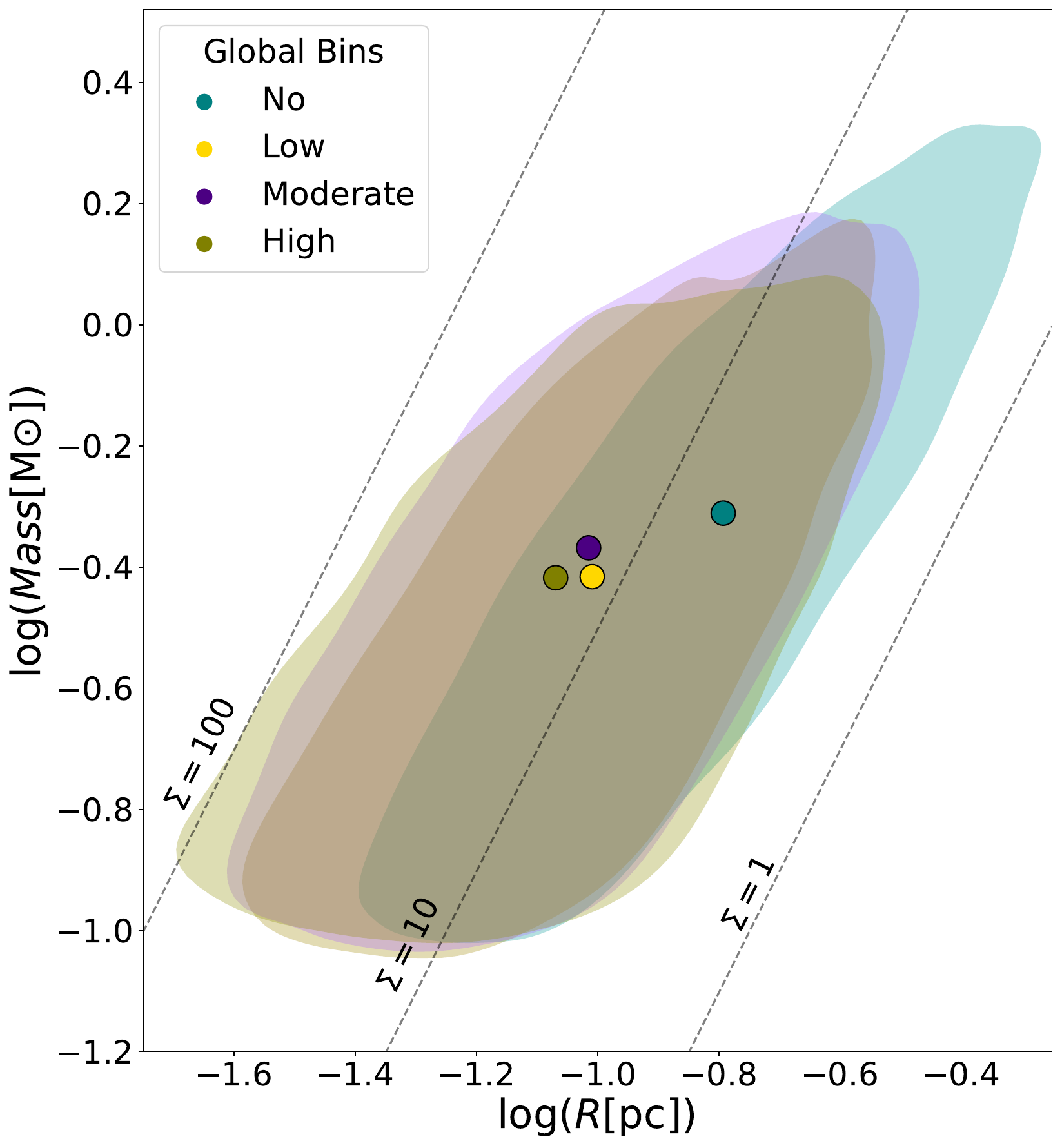}
    \includegraphics[width = .47 \textwidth, keepaspectratio]{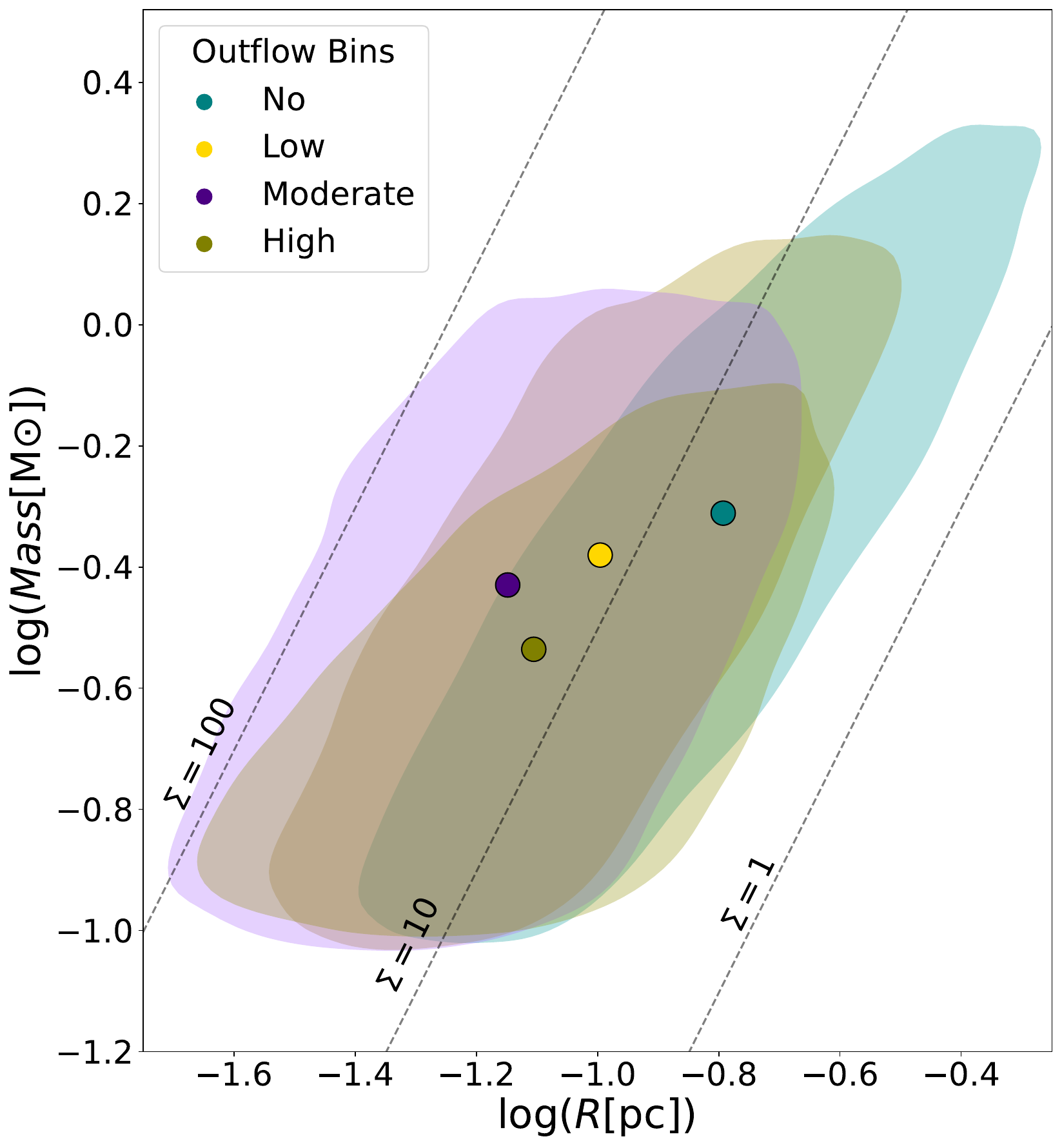}
    \includegraphics[width = .47 \textwidth, keepaspectratio]{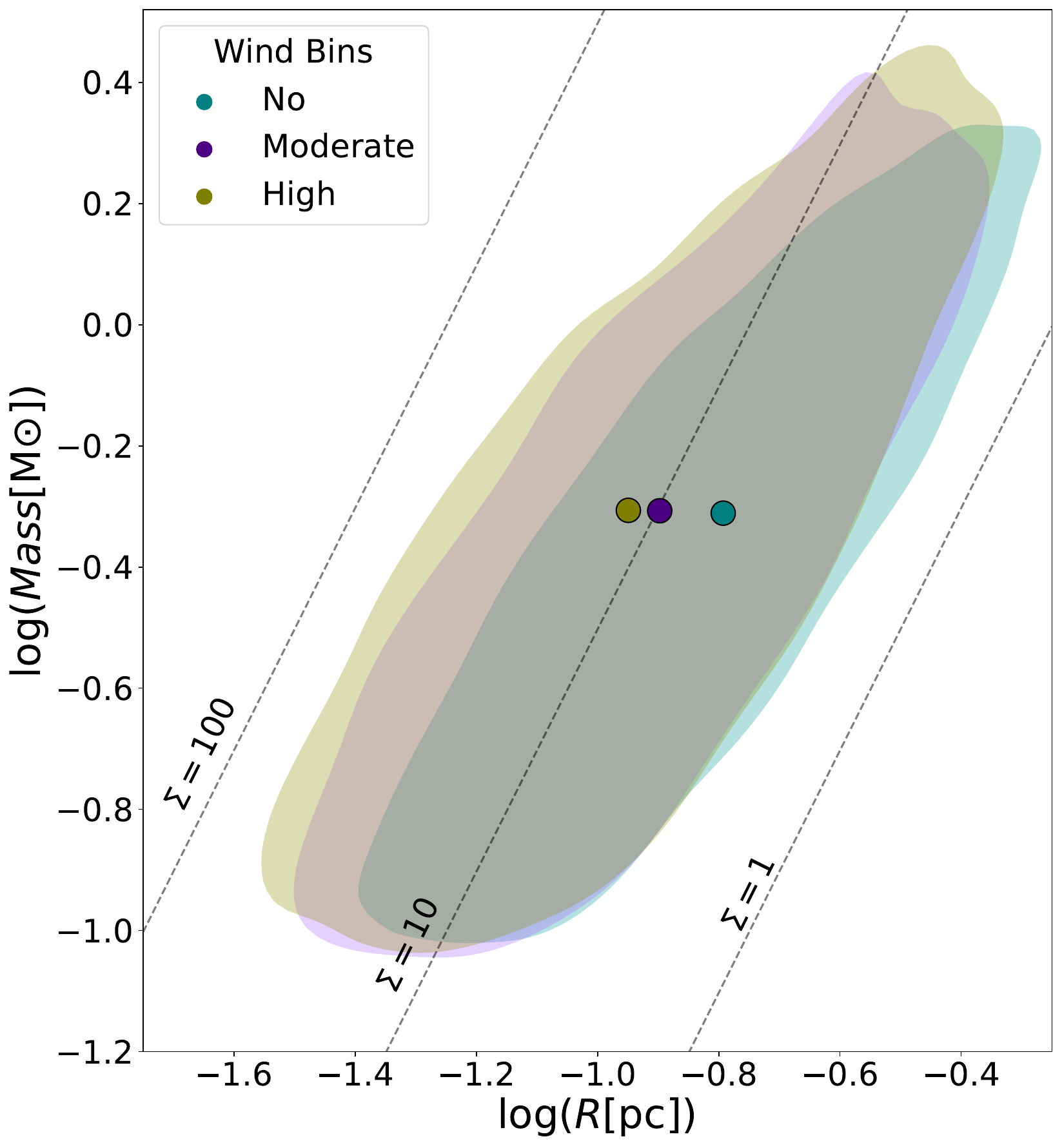}
    \includegraphics[width = .47 \textwidth, keepaspectratio]{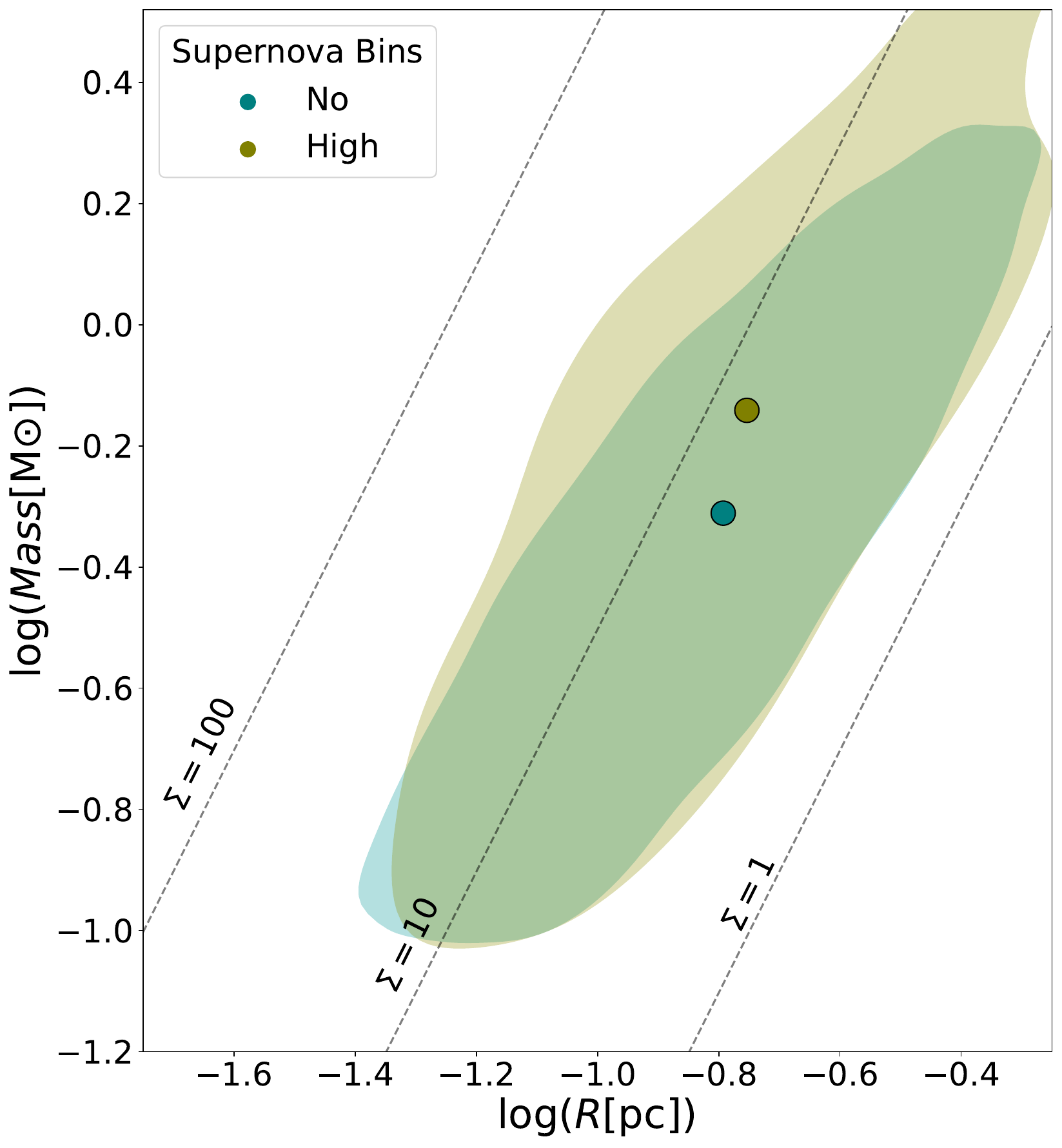}
    \caption{Scaling relation between mass and size ($M$ vs $R$) of cores in the outflow bins. The dashed lines represent the constant surface mass densities in $\rm{M}_\odot \rm{pc}^{-2}$. The symbols and conventions for the subplots follow Fig. \ref{fig: m2e4a2 larson feedback}.}
    \label{fig: m2e4a2 size mass feedback}
\end{figure*}

Fig. \ref{fig: m2e4a2 size mass feedback} compares the mass and size of the cores. We observe trends similar to the above scaling relations, with stellar feedback leading to smaller cores. This effect is most distinct for outflows. 
However, the differences in various distributions are not as significant as in other scaling relations. The similar average mass and radius of the cores indicate that core selection is stable despite environmental variations, and the differences in distributions are due to stellar feedback.
The slopes of our mass-size ellipses are $\sim$ 1-1.5 (Table \ref{tab: size mass slopes and width}). They are different from the slopes of the dense cores observed in the Galaxy, which are typically $\sim$2 or higher \citep{lada2008ApJ...672..410L}.

\begin{table}[]
\caption{Size-mass relation: slopes and widths of PCA ellipse.}
\label{tab: size mass slopes and width}
\centering
\begin{tabular}{lllll}
\hline
Feedback Bins & Slope & $\sigma_{\rm{slope}}$ & Scatter & $\sigma_{\rm{scatter}}$ \\ \hline
No            & 1.29  & < 0.01                  & 0.34    & < 0.01                    \\ \hline
\multicolumn{5}{c}{Global}       \\
Low           & 1.28  & 0.01                  & 0.4     & < 0.01                    \\
Moderate      & 1.25  & 0.01                  & 0.48    & < 0.01                    \\
High          & 1.16  & 0.01                  & 0.52    & < 0.01                    \\ \hline
\multicolumn{5}{c}{Outflow}       \\
Low           & 1.28  & < 0.01                  & 0.39    & < 0.01                    \\
Moderate      & 1.22  & 0.01                  & 0.51    & < 0.01                    \\
High          & 0.87  & 0.01                  & 0.49    & < 0.01                    \\ \hline
\multicolumn{5}{c}{Wind}       \\
Moderate      & 1.35  & 0.02                  & 0.4     & < 0.01                    \\
High          & 1.25  & 0.01                  & 0.46    & < 0.01                    \\ \hline
\multicolumn{5}{c}{Supernova}       \\
High          & 1.54  & 0.04                  & 0.38    & 0.01                    \\ \hline
\end{tabular}
\tablefoot{The lines of constant surface mass densities in Fig. \ref{fig: m2e4a2 size mass feedback} represent a slope of 2. The symbols and conventions follow Table \ref{tab: larson slopes and width}.}
\end{table}

\section{Summary and conclusions}\label{sec: summary}

In this paper we explore the effects of different stellar feedback mechanisms on molecular cores.
We performed a dendrogram analysis on the { gas density modeled in a 20,000} M$_\odot$ \starforge simulation. We then analysed the properties of the dendrogram leaves or cores to understand how these structures are affected by protostellar outflows, stellar winds, and supernova.

Our main conclusions are the following.
\begin{enumerate}

\item{Cores strongly affected by feedback have a higher velocity dispersion on average than cores with less feedback, especially when comparing similar-sized structures. This is observed for all three types of feedback mechanisms.
We attribute this to the injection of momentum into the dense gas by these feedback mechanisms.}


\item{Cores influenced by feedback tend to be smaller and denser on average compared to the pristine (no-feedback bin) cores. This could be the result of 
compression of the molecular gas as a result of stellar feedback or because these cores are more evolved and are gravitationally collasping. There is no significant change in the radius of the cores from low- to high-feedback bins, however, the high-supernova bin has the largest cores on average.}

\item{Feedback doesn't affect core mass 
significantly. The slight changes in the mass distributions that occur for different amounts of feedback 
follow the same trends as core radius.}

\item{The amount of feedback influence correlates with the virial parameter, where cores that are characterized by the presence of more feedback are more gravitationally unbound. We see this trend for all types of feedbacks.}

\item{The scaling relations suggest a scenario where outflow and wind feedback 
leads to smaller cores whereas supernova 
feedback tends to increase core size.
Outflows likely disperse molecular gas, whereas winds compress the cores and disperse the gas. The shocks produced 
by the supernova explosion leads to large cores.}

\item{The cores with less feedback adhere more closely to the traditional star formation scaling relations (e.g., Larson's laws), evidenced by their reduced scatter.}

\end{enumerate}

\begin{acknowledgement}
The authors thank the anonymous referee for a thorough and constructive report which has improved the quality of the manuscript. This research made use of astrodendro, a Python package to compute dendrograms of Astronomical data (http://www.dendrograms.org/). Support for MYG was provided by NASA through the NASA Hubble Fellowship grant \#HST-HF2-51479 awarded  by  the  Space  Telescope  Science  Institute,  which  is  operated  by  the   Association  of  Universities  for  Research  in  Astronomy,  Inc.,  for  NASA,  under  contract NAS5-26555. This research is part of the Frontera computing project at the Texas Advanced Computing Center. Frontera is made possible by National Science Foundation award OAC-1818253.
\end{acknowledgement}

\footnotesize{
\bibliographystyle{aa}
\bibliography{reference}

\begin{thebibliography}{85}
\expandafter\ifx\csname natexlab\endcsname\relax\def\natexlab#1{#1}\fi

\bibitem[{{Agertz} \& {Kravtsov}(2015)}]{agertz2015}
{Agertz}, O. \& {Kravtsov}, A.~V. 2015, \apj, 804, 18

\bibitem[{{Arce} {et~al.}(2011){Arce}, {Borkin}, {Goodman}, {Pineda}, \& {Beaumont}}]{arce2011ApJ...742..105A}
{Arce}, H.~G., {Borkin}, M.~A., {Goodman}, A.~A., {Pineda}, J.~E., \& {Beaumont}, C.~N. 2011, \apj, 742, 105

\bibitem[{{Arce} \& {Goodman}(2001)}]{arce2001ApJ...554..132A}
{Arce}, H.~G. \& {Goodman}, A.~A. 2001, \apj, 554, 132

\bibitem[{{Arce} {et~al.}(2007){Arce}, {Shepherd}, {Gueth}, {Lee}, {Bachiller}, {Rosen}, \& {Beuther}}]{arce2007prpl.conf..245A}
{Arce}, H.~G., {Shepherd}, D., {Gueth}, F., {et~al.} 2007, in Protostars and Planets V, ed. B.~{Reipurth}, D.~{Jewitt}, \& K.~{Keil}, 245

\bibitem[{{Ballesteros-Paredes} {et~al.}(2020){Ballesteros-Paredes}, {Andr{\'e}}, {Hennebelle}, {Klessen}, {Kruijssen}, {Chevance}, {Nakamura}, {Adamo}, \& {V{\'a}zquez-Semadeni}}]{ballesteros-paredes2020}
{Ballesteros-Paredes}, J., {Andr{\'e}}, P., {Hennebelle}, P., {et~al.} 2020, \ssr, 216, 76

\bibitem[{{Bally}(2016)}]{bally2016}
{Bally}, J. 2016, \araa, 54, 491

\bibitem[{{Beaumont} {et~al.}(2013){Beaumont}, {Offner}, {Shetty}, {Glover}, \& {Goodman}}]{beaumont2013ApJ...777..173B}
{Beaumont}, C.~N., {Offner}, S. S.~R., {Shetty}, R., {Glover}, S. C.~O., \& {Goodman}, A.~A. 2013, \apj, 777, 173

\bibitem[{{Bieri} {et~al.}(2023){Bieri}, {Naab}, {Geen}, {Coles}, {Pakmor}, \& {Walch}}]{bieri2023MNRAS.523.6336B}
{Bieri}, R., {Naab}, T., {Geen}, S., {et~al.} 2023, \mnras, 523, 6336

\bibitem[{{Blitz}(1993)}]{blitz1993prpl.conf..125B}
{Blitz}, L. 1993, in Protostars and Planets III, ed. E.~H. {Levy} \& J.~I. {Lunine}, 125

\bibitem[{{Chen} {et~al.}(2019){Chen}, {Pineda}, {Goodman}, {Burkert}, {Offner}, {Friesen}, {Myers}, {Alves}, {Arce}, {Caselli}, {Chac{\'o}n-Tanarro}, {Chen}, {Di Francesco}, {Ginsburg}, {Keown}, {Kirk}, {Martin}, {Matzner}, {Punanova}, {Redaelli}, {Rosolowsky}, {Scibelli}, {Seo}, {Shirley}, {Singh}, \& {GAS Collaboration}}]{chen2019}
{Chen}, H. H.-H., {Pineda}, J.~E., {Goodman}, A.~A., {et~al.} 2019, \apj, 877, 93

\bibitem[{{Chevance} {et~al.}(2023){Chevance}, {Krumholz}, {McLeod}, {Ostriker}, {Rosolowsky}, \& {Sternberg}}]{chevance2023}
{Chevance}, M., {Krumholz}, M.~R., {McLeod}, A.~F., {et~al.} 2023, in Astronomical Society of the Pacific Conference Series, Vol. 534, Protostars and Planets VII, ed. S.~{Inutsuka}, Y.~{Aikawa}, T.~{Muto}, K.~{Tomida}, \& M.~{Tamura}, 1

\bibitem[{{Colombo} {et~al.}(2019){Colombo}, {Rosolowsky}, {Duarte-Cabral}, {Ginsburg}, {Glenn}, {Zetterlund}, {Hernandez}, {Dempsey}, \& {Currie}}]{colombo2019}
{Colombo}, D., {Rosolowsky}, E., {Duarte-Cabral}, A., {et~al.} 2019, \mnras, 483, 4291

\bibitem[{{Cunningham} {et~al.}(2011){Cunningham}, {Klein}, {Krumholz}, \& {McKee}}]{cunningham2011ApJ...740..107C}
{Cunningham}, A.~J., {Klein}, R.~I., {Krumholz}, M.~R., \& {McKee}, C.~F. 2011, \apj, 740, 107

\bibitem[{{Davis} {et~al.}(2008){Davis}, {Scholz}, {Lucas}, {Smith}, \& {Adamson}}]{davis2008MNRAS.387..954D}
{Davis}, C.~J., {Scholz}, P., {Lucas}, P., {Smith}, M.~D., \& {Adamson}, A. 2008, \mnras, 387, 954

\bibitem[{{Deharveng} {et~al.}(2010){Deharveng}, {Schuller}, {Anderson}, {Zavagno}, {Wyrowski}, {Menten}, {Bronfman}, {Testi}, {Walmsley}, \& {Wienen}}]{deharveng2010A&A...523A...6D}
{Deharveng}, L., {Schuller}, F., {Anderson}, L.~D., {et~al.} 2010, \aap, 523, A6

\bibitem[{{Dokara} {et~al.}(2023){Dokara}, {Gong}, {Reich}, {Rugel}, {Brunthaler}, {Menten}, {Cotton}, {Dzib}, {Khan}, {Medina}, {Nguyen}, {Ortiz-Le{\'o}n}, {Urquhart}, {Wyrowski}, {Yang}, {Anderson}, {Beuther}, {Csengeri}, {M{\"u}ller}, {Ott}, {Pandian}, \& {Roy}}]{dokara2023A&A...671A.145D}
{Dokara}, R., {Gong}, Y., {Reich}, W., {et~al.} 2023, \aap, 671, A145

\bibitem[{{Duarte-Cabral} {et~al.}(2012){Duarte-Cabral}, {Chrysostomou}, {Peretto}, {Fuller}, {Matthews}, {Schieven}, \& {Davis}}]{duarte-cabral2012A&A...543A.140D}
{Duarte-Cabral}, A., {Chrysostomou}, A., {Peretto}, N., {et~al.} 2012, \aap, 543, A140

\bibitem[{{Duarte-Cabral} {et~al.}(2021){Duarte-Cabral}, {Colombo}, {Urquhart}, {Ginsburg}, {Russeil}, {Schuller}, {Anderson}, {Barnes}, {Beltr{\'a}n}, {Beuther}, {Bontemps}, {Bronfman}, {Csengeri}, {Dobbs}, {Eden}, {Giannetti}, {Kauffmann}, {Mattern}, {Medina}, {Menten}, {Lee}, {Pettitt}, {Riener}, {Rigby}, {Traficante}, {Veena}, {Wienen}, {Wyrowski}, {Agurto}, {Azagra}, {Cesaroni}, {Finger}, {Gonzalez}, {Henning}, {Hernandez}, {Kainulainen}, {Leurini}, {Lopez}, {Mac-Auliffe}, {Mazumdar}, {Molinari}, {Motte}, {Muller}, {Nguyen-Luong}, {Parra}, {Perez-Beaupuits}, {Montenegro-Montes}, {Moore}, {Ragan}, {S{\'a}nchez-Monge}, {Sanna}, {Schilke}, {Schisano}, {Schneider}, {Suri}, {Testi}, {Torstensson}, {Venegas}, {Wang}, \& {Zavagno}}]{duarte_cabral2021}
{Duarte-Cabral}, A., {Colombo}, D., {Urquhart}, J.~S., {et~al.} 2021, \mnras, 500, 3027

\bibitem[{{Duarte-Cabral} \& {Dobbs}(2016)}]{duarte-cabral2016MNRAS.458.3667D}
{Duarte-Cabral}, A. \& {Dobbs}, C.~L. 2016, \mnras, 458, 3667

\bibitem[{{Dubner} \& {Giacani}(2015)}]{dubner2015A&ARv..23....3D}
{Dubner}, G. \& {Giacani}, E. 2015, \aapr, 23, 3

\bibitem[{{Federrath} {et~al.}(2014){Federrath}, {Schr{\"o}n}, {Banerjee}, \& {Klessen}}]{federrath2014ApJ...790..128F}
{Federrath}, C., {Schr{\"o}n}, M., {Banerjee}, R., \& {Klessen}, R.~S. 2014, \apj, 790, 128

\bibitem[{{Fendt} \& {{\v{C}}emelji{\'c}}(2002)}]{fendt2002A&A...395.1045F}
{Fendt}, C. \& {{\v{C}}emelji{\'c}}, M. 2002, \aap, 395, 1045

\bibitem[{{Fichtner} {et~al.}(2022){Fichtner}, {Grassitelli}, {Romano-D{\'\i}az}, \& {Porciani}}]{fichtner2022MNRAS.512.4573F}
{Fichtner}, Y.~A., {Grassitelli}, L., {Romano-D{\'\i}az}, E., \& {Porciani}, C. 2022, \mnras, 512, 4573

\bibitem[{{Frank} {et~al.}(2014){Frank}, {Ray}, {Cabrit}, {Hartigan}, {Arce}, {Bacciotti}, {Bally}, {Benisty}, {Eisl{\"o}ffel}, {G{\"u}del}, {Lebedev}, {Nisini}, \& {Raga}}]{frank2014prpl.conf..451F}
{Frank}, A., {Ray}, T.~P., {Cabrit}, S., {et~al.} 2014, in Protostars and Planets VI, ed. H.~{Beuther}, R.~S. {Klessen}, C.~P. {Dullemond}, \& T.~{Henning}, 451--474

\bibitem[{{Friesen} {et~al.}(2016){Friesen}, {Bourke}, {Di Francesco}, {Gutermuth}, \& {Myers}}]{friesen2016ApJ...833..204F}
{Friesen}, R.~K., {Bourke}, T.~L., {Di Francesco}, J., {Gutermuth}, R., \& {Myers}, P.~C. 2016, \apj, 833, 204

\bibitem[{{Fuente} {et~al.}(2002){Fuente}, {Mart{\i}n-Pintado}, {Bachiller}, {Rodr{\i}guez-Franco}, \& {Palla}}]{fuente2002A&A...387..977F}
{Fuente}, A., {Mart{\i}n-Pintado}, J., {Bachiller}, R., {Rodr{\i}guez-Franco}, A., \& {Palla}, F. 2002, \aap, 387, 977

\bibitem[{{Gatto} {et~al.}(2017){Gatto}, {Walch}, {Naab}, {Girichidis}, {W{\"u}nsch}, {Glover}, {Klessen}, {Clark}, {Peters}, {Derigs}, {Baczynski}, \& {Puls}}]{gatto2017}
{Gatto}, A., {Walch}, S., {Naab}, T., {et~al.} 2017, \mnras, 466, 1903

\bibitem[{{Geen} {et~al.}(2023){Geen}, {Agrawal}, {Crowther}, {Keller}, {de Koter}, {Keszthelyi}, {van de Voort}, {Ali}, {Backs}, {Bonne}, {Brugaletta}, {Derkink}, {Ekstr{\"o}m}, {Fichtner}, {Grassitelli}, {G{\"o}tberg}, {Higgins}, {Laplace}, {You Liow}, {Lorenzo}, {McLeod}, {Meynet}, {Newsome}, {Andr{\'e} Oliva}, {Ramachandran}, {Rey}, {Rieder}, {Romano-D{\'\i}az}, {Sabhahit}, {Sander}, {Sarwar}, {Stinshoff}, {Stoop}, {Sz{\'e}csi}, {Trebitsch}, {Vink}, \& {Winch}}]{geen2023}
{Geen}, S., {Agrawal}, P., {Crowther}, P.~A., {et~al.} 2023, arXiv e-prints, arXiv:2301.13611

\bibitem[{{Geen} {et~al.}(2015){Geen}, {Rosdahl}, {Blaizot}, {Devriendt}, \& {Slyz}}]{geen2015MNRAS.448.3248G}
{Geen}, S., {Rosdahl}, J., {Blaizot}, J., {Devriendt}, J., \& {Slyz}, A. 2015, \mnras, 448, 3248

\bibitem[{{Girichidis} {et~al.}(2020){Girichidis}, {Offner}, {Kritsuk}, {Klessen}, {Hennebelle}, {Kruijssen}, {Krause}, {Glover}, \& {Padovani}}]{girischidis2020SSRv..216...68G}
{Girichidis}, P., {Offner}, S. S.~R., {Kritsuk}, A.~G., {et~al.} 2020, \ssr, 216, 68

\bibitem[{{Grishunin} {et~al.}(2024){Grishunin}, {Weiss}, {Colombo}, {Chevance}, {Chen}, {G{\"u}sten}, {Rubio}, {Hunt}, {Wyrowski}, {Harrington}, {Menten}, \& {Herrera-Camus}}]{grishunin2024A&A...682A.137G}
{Grishunin}, K., {Weiss}, A., {Colombo}, D., {et~al.} 2024, \aap, 682, A137

\bibitem[{{Grudi{\'c}} \& {Gurvich}(2021)}]{pytreegrav}
{Grudi{\'c}}, M. \& {Gurvich}, A. 2021, The Journal of Open Source Software, 6, 3675

\bibitem[{{Grudi{\'c}} {et~al.}(2021){Grudi{\'c}}, {Guszejnov}, {Hopkins}, {Offner}, \& {Faucher-Gigu{\`e}re}}]{grudic2021}
{Grudi{\'c}}, M.~Y., {Guszejnov}, D., {Hopkins}, P.~F., {Offner}, S. S.~R., \& {Faucher-Gigu{\`e}re}, C.-A. 2021, \mnras, 506, 2199

\bibitem[{{Grudi{\'c}} {et~al.}(2022){Grudi{\'c}}, {Guszejnov}, {Offner}, {Rosen}, {Raju}, {Faucher-Gigu{\`e}re}, \& {Hopkins}}]{grudic2022}
{Grudi{\'c}}, M.~Y., {Guszejnov}, D., {Offner}, S. S.~R., {et~al.} 2022, \mnras, 512, 216

\bibitem[{{Guszejnov} {et~al.}(2021){Guszejnov}, {Grudi{\'c}}, {Hopkins}, {Offner}, \& {Faucher-Gigu{\`e}re}}]{guszejnov2021MNRAS.502.3646G}
{Guszejnov}, D., {Grudi{\'c}}, M.~Y., {Hopkins}, P.~F., {Offner}, S. S.~R., \& {Faucher-Gigu{\`e}re}, C.-A. 2021, \mnras, 502, 3646

\bibitem[{{Guszejnov} {et~al.}(2022){Guszejnov}, {Grudi{\'c}}, {Offner}, {Faucher-Gigu{\`e}re}, {Hopkins}, \& {Rosen}}]{guszejnov2022}
{Guszejnov}, D., {Grudi{\'c}}, M.~Y., {Offner}, S. S.~R., {et~al.} 2022, \mnras, 515, 4929

\bibitem[{{Heyer} {et~al.}(2009){Heyer}, {Krawczyk}, {Duval}, \& {Jackson}}]{heyer2009}
{Heyer}, M., {Krawczyk}, C., {Duval}, J., \& {Jackson}, J.~M. 2009, \apj, 699, 1092

\bibitem[{{Hopkins}(2015)}]{gizmo}
{Hopkins}, P.~F. 2015, \mnras, 450, 53

\bibitem[{{Hopkins} {et~al.}(2012){Hopkins}, {Quataert}, \& {Murray}}]{hopkins2012}
{Hopkins}, P.~F., {Quataert}, E., \& {Murray}, N. 2012, \mnras, 421, 3488

\bibitem[{{Hopkins} \& {Raives}(2016)}]{hopkins2016MNRAS.455...51H}
{Hopkins}, P.~F. \& {Raives}, M.~J. 2016, \mnras, 455, 51

\bibitem[{{Kauffmann} {et~al.}(2013){Kauffmann}, {Pillai}, \& {Goldsmith}}]{kauffmann2013ApJ...779..185K}
{Kauffmann}, J., {Pillai}, T., \& {Goldsmith}, P.~F. 2013, \apj, 779, 185

\bibitem[{{Keown} {et~al.}(2017){Keown}, {Di Francesco}, {Kirk}, {Friesen}, {Pineda}, {Rosolowsky}, {Ginsburg}, {Offner}, {Caselli}, {Alves}, {Chac{\'o}n-Tanarro}, {Punanova}, {Redaelli}, {Seo}, {Matzner}, {Chun-Yuan Chen}, {Goodman}, {Chen}, {Shirley}, {Singh}, {Arce}, {Martin}, \& {Myers}}]{keown2017ApJ...850....3K}
{Keown}, J., {Di Francesco}, J., {Kirk}, H., {et~al.} 2017, \apj, 850, 3

\bibitem[{{Kirchschlager} {et~al.}(2024){Kirchschlager}, {Sartorio}, {De Looze}, {Barlow}, {Schmidt}, \& {Priestley}}]{kirchschlager2024MNRAS.tmp..359K}
{Kirchschlager}, F., {Sartorio}, N.~S., {De Looze}, I., {et~al.} 2024, \mnras [\eprint[arXiv]{2402.00701}]

\bibitem[{{Kirsanova} \& {Pavlyuchenkov}(2023)}]{kirsanova2023IAUS..362..268K}
{Kirsanova}, M.~S. \& {Pavlyuchenkov}, Y.~N. 2023, in The Predictive Power of Computational Astrophysics as a Discover Tool, ed. D.~{Bisikalo}, D.~{Wiebe}, \& C.~{Boily}, Vol. 362, 268--272

\bibitem[{{Kramer} {et~al.}(2023){Kramer}, {Adam}, {Ade}, {Ajeddig}, {Andre}, {Artis}, {Aussel}, {Beelen}, {Beno}, {Berta}, {Bing}, {Bourrion}, {Calvo}, {Caselli}, {Catalano}, {DePetris}, {Desert}, {Doyle}, {Driessen}, {Ejlali}, {Fuente}, {Gomez}, {Goupy}, {Hanser}, {Katsioli}, {Keruzore}, {Ladjelate}, {Lagache}, {Leclercq}, {Lestrade}, {Macias-Perez}, {Madden}, {Maury}, {Mauskopf}, {Mayet}, {Monfardini}, {Moyer-Anin}, {Munoz-Echeverria}, {Navarro-Almaida}, {Perotto}, {Pisano}, {Ponthieu}, {Reveret}, {Rigby}, {Ritacco}, {Romero}, {Roussel}, {Ruppin}, {Schuster}, {Sievers}, {Tucker}, \& {Zylka}}]{kramer2023arXiv231001044K}
{Kramer}, C., {Adam}, R., {Ade}, P., {et~al.} 2023, arXiv e-prints, arXiv:2310.01044

\bibitem[{{Kruijssen} {et~al.}(2019){Kruijssen}, {Schruba}, {Chevance}, {Longmore}, {Hygate}, {Haydon}, {McLeod}, {Dalcanton}, {Tacconi}, \& {van Dishoeck}}]{kruijssen2019}
{Kruijssen}, J.~M.~D., {Schruba}, A., {Chevance}, M., {et~al.} 2019, \nat, 569, 519

\bibitem[{{Krumholz} {et~al.}(2014){Krumholz}, {Bate}, {Arce}, {Dale}, {Gutermuth}, {Klein}, {Li}, {Nakamura}, \& {Zhang}}]{krumholz2014}
{Krumholz}, M.~R., {Bate}, M.~R., {Arce}, H.~G., {et~al.} 2014, in Protostars and Planets VI, ed. H.~{Beuther}, R.~S. {Klessen}, C.~P. {Dullemond}, \& T.~{Henning}, 243--266

\bibitem[{{Lada} {et~al.}(2008){Lada}, {Muench}, {Rathborne}, {Alves}, \& {Lombardi}}]{lada2008ApJ...672..410L}
{Lada}, C.~J., {Muench}, A.~A., {Rathborne}, J., {Alves}, J.~F., \& {Lombardi}, M. 2008, \apj, 672, 410

\bibitem[{{Larson}(1981)}]{larson1981}
{Larson}, R.~B. 1981, \mnras, 194, 809

\bibitem[{{Lee} {et~al.}(2002){Lee}, {Mundy}, {Stone}, \& {Ostriker}}]{lee2002ApJ...576..294L}
{Lee}, C.-F., {Mundy}, L.~G., {Stone}, J.~M., \& {Ostriker}, E.~C. 2002, \apj, 576, 294

\bibitem[{{Mairs} {et~al.}(2014){Mairs}, {Johnstone}, {Offner}, \& {Schnee}}]{mairs2014ApJ...783...60M}
{Mairs}, S., {Johnstone}, D., {Offner}, S. S.~R., \& {Schnee}, S. 2014, \apj, 783, 60

\bibitem[{{Maret} {et~al.}(2009){Maret}, {Bergin}, {Neufeld}, {Green}, {Watson}, {Harwit}, {Kristensen}, {Melnick}, {Sonnentrucker}, {Tolls}, {Werner}, {Willacy}, \& {Yuan}}]{maret2009ApJ...698.1244M}
{Maret}, S., {Bergin}, E.~A., {Neufeld}, D.~A., {et~al.} 2009, \apj, 698, 1244

\bibitem[{{McKee} \& {Ostriker}(1977)}]{mckee1977ApJ...218..148M}
{McKee}, C.~F. \& {Ostriker}, J.~P. 1977, \apj, 218, 148

\bibitem[{{Myers}(1983)}]{myers1983}
{Myers}, P.~C. 1983, \apj, 270, 105

\bibitem[{{Narayanan} {et~al.}(2012){Narayanan}, {Snell}, \& {Bemis}}]{narayanan2012MNRAS.425.2641N}
{Narayanan}, G., {Snell}, R., \& {Bemis}, A. 2012, \mnras, 425, 2641

\bibitem[{{Neralwar} {et~al.}(2022){Neralwar}, {Colombo}, {Duarte-Cabral}, {Urquhart}, {Mattern}, {Wyrowski}, {Menten}, {Barnes}, {S{\'a}nchez-Monge}, {Rigby}, {Mazumdar}, {Eden}, {Csengeri}, {Dobbs}, {Veena}, {Neupane}, {Henning}, {Schuller}, {Leurini}, {Wienen}, {Yang}, {Ragan}, {Medina}, \& {Nguyen-Luong}}]{neralwar2022b}
{Neralwar}, K.~R., {Colombo}, D., {Duarte-Cabral}, A., {et~al.} 2022, \aap, 664, A84

\bibitem[{{Nony} {et~al.}(2023){Nony}, {Galv{\'a}n-Madrid}, {Motte}, {Pouteau}, {Cunningham}, {Louvet}, {Stutz}, {Lefloch}, {Bontemps}, {Brouillet}, {Ginsburg}, {Joncour}, {Herpin}, {Sanhueza}, {Csengeri}, {Towner}, {Bonfand}, {Fern{\'a}ndez-L{\'o}pez}, {Baug}, {Bronfman}, {Busquet}, {Di Francesco}, {Gusdorf}, {Lu}, {Olguin}, {Valeille-Manet}, \& {Whitworth}}]{nony2023A&A...674A..75N}
{Nony}, T., {Galv{\'a}n-Madrid}, R., {Motte}, F., {et~al.} 2023, \aap, 674, A75

\bibitem[{{Offner} \& {Arce}(2015)}]{offner2015ApJ...811..146O}
{Offner}, S. S.~R. \& {Arce}, H.~G. 2015, \apj, 811, 146

\bibitem[{{Offner} \& {Chaban}(2017)}]{offner2017}
{Offner}, S. S.~R. \& {Chaban}, J. 2017, \apj, 847, 104

\bibitem[{{Offner} {et~al.}(2011){Offner}, {Lee}, {Goodman}, \& {Arce}}]{Offner2011}
{Offner}, S. S.~R., {Lee}, E.~J., {Goodman}, A.~A., \& {Arce}, H. 2011, \apj, 743, 91

\bibitem[{{Offner} \& {Liu}(2018)}]{offner2018}
{Offner}, S. S.~R. \& {Liu}, Y. 2018, Nature Astronomy, 2, 896

\bibitem[{{Offner} {et~al.}(2022){Offner}, {Taylor}, {Markey}, {Chen}, {Pineda}, {Goodman}, {Burkert}, {Ginsburg}, \& {Choudhury}}]{offner2022MNRAS.517..885O}
{Offner}, S. S.~R., {Taylor}, J., {Markey}, C., {et~al.} 2022, \mnras, 517, 885

\bibitem[{{Olmi} {et~al.}(2023){Olmi}, {Brand}, \& {Elia}}]{olmi2023MNRAS.518.1917O}
{Olmi}, L., {Brand}, J., \& {Elia}, D. 2023, \mnras, 518, 1917

\bibitem[{{O'Neill} {et~al.}(2021){O'Neill}, {Cosentino}, {Tan}, {Cheng}, \& {Liu}}]{oneil2021AAS...23711203O}
{O'Neill}, T., {Cosentino}, G., {Tan}, J.~C., {Cheng}, Y., \& {Liu}, M. 2021, in American Astronomical Society Meeting Abstracts, Vol.~53, American Astronomical Society Meeting Abstracts, 112.03

\bibitem[{{Pabst} {et~al.}(2019){Pabst}, {Higgins}, {Goicoechea}, {Teyssier}, {Berne}, {Chambers}, {Wolfire}, {Suri}, {Guesten}, {Stutzki}, {Graf}, {Risacher}, \& {Tielens}}]{pabst2019Natur.565..618P}
{Pabst}, C., {Higgins}, R., {Goicoechea}, J.~R., {et~al.} 2019, \nat, 565, 618

\bibitem[{{Pabst} {et~al.}(2020){Pabst}, {Goicoechea}, {Teyssier}, {Bern{\'e}}, {Higgins}, {Chambers}, {Kabanovic}, {G{\"u}sten}, {Stutzki}, \& {Tielens}}]{pabst2020A&A...639A...2P}
{Pabst}, C.~H.~M., {Goicoechea}, J.~R., {Teyssier}, D., {et~al.} 2020, \aap, 639, A2

\bibitem[{{Peters} {et~al.}(2017){Peters}, {Naab}, {Walch}, {Glover}, {Girichidis}, {Pellegrini}, {Klessen}, {W{\"u}nsch}, {Gatto}, \& {Baczynski}}]{peters2017MNRAS.466.3293P}
{Peters}, T., {Naab}, T., {Walch}, S., {et~al.} 2017, \mnras, 466, 3293

\bibitem[{{Rosolowsky} {et~al.}(2021){Rosolowsky}, {Hughes}, {Leroy}, {Sun}, {Querejeta}, {Schruba}, {Usero}, {Herrera}, {Liu}, {Pety}, {Saito}, {Be{\v{s}}li{\'c}}, {Bigiel}, {Blanc}, {Chevance}, {Dale}, {Deger}, {Faesi}, {Glover}, {Henshaw}, {Klessen}, {Kruijssen}, {Larson}, {Lee}, {Meidt}, {Mok}, {Schinnerer}, {Thilker}, \& {Williams}}]{rosolowsky2021MNRAS.502.1218R}
{Rosolowsky}, E., {Hughes}, A., {Leroy}, A.~K., {et~al.} 2021, \mnras, 502, 1218

\bibitem[{{Rosolowsky} {et~al.}(2008){Rosolowsky}, {Pineda}, {Kauffmann}, \& {Goodman}}]{rosolowsky2008ApJ...679.1338R}
{Rosolowsky}, E.~W., {Pineda}, J.~E., {Kauffmann}, J., \& {Goodman}, A.~A. 2008, \apj, 679, 1338

\bibitem[{{Schneider} {et~al.}(2020){Schneider}, {Simon}, {Guevara}, {Buchbender}, {Higgins}, {Okada}, {Stutzki}, {G{\"u}sten}, {Anderson}, {Bally}, {Beuther}, {Bonne}, {Bontemps}, {Chambers}, {Csengeri}, {Graf}, {Gusdorf}, {Jacobs}, {Justen}, {Kabanovic}, {Karim}, {Luisi}, {Menten}, {Mertens}, {Mookerjea}, {Ossenkopf-Okada}, {Pabst}, {Pound}, {Richter}, {Reyes}, {Ricken}, {R{\"o}llig}, {Russeil}, {S{\'a}nchez-Monge}, {Sandell}, {Tiwari}, {Wiesemeyer}, {Wolfire}, {Wyrowski}, {Zavagno}, \& {Tielens}}]{schneider2020PASP..132j4301S}
{Schneider}, N., {Simon}, R., {Guevara}, C., {et~al.} 2020, \pasp, 132, 104301

\bibitem[{{Schuller} {et~al.}(2009){Schuller}, {Menten}, {Contreras}, {Wyrowski}, {Schilke}, {Bronfman}, {Henning}, {Walmsley}, {Beuther}, {Bontemps}, {Cesaroni}, {Deharveng}, {Garay}, {Herpin}, {Lefloch}, {Linz}, {Mardones}, {Minier}, {Molinari}, {Motte}, {Nyman}, {Reveret}, {Risacher}, {Russeil}, {Schneider}, {Testi}, {Troost}, {Vasyunina}, {Wienen}, {Zavagno}, {Kovacs}, {Kreysa}, {Siringo}, \& {Wei{\ss}}}]{schuller2009A&A...504..415S}
{Schuller}, F., {Menten}, K.~M., {Contreras}, Y., {et~al.} 2009, \aap, 504, 415

\bibitem[{{Seo} {et~al.}(2015){Seo}, {Shirley}, {Goldsmith}, {Ward-Thompson}, {Kirk}, {Schmalzl}, {Lee}, {Friesen}, {Langston}, {Masters}, \& {Garwood}}]{seo2015ApJ...805..185S}
{Seo}, Y.~M., {Shirley}, Y.~L., {Goldsmith}, P., {et~al.} 2015, \apj, 805, 185

\bibitem[{{Smullen} {et~al.}(2020){Smullen}, {Kratter}, {Offner}, {Lee}, \& {Chen}}]{smullen2020MNRAS.497.4517S}
{Smullen}, R.~A., {Kratter}, K.~M., {Offner}, S. S.~R., {Lee}, A.~T., \& {Chen}, H. H.-H. 2020, \mnras, 497, 4517

\bibitem[{{Sokol} {et~al.}(2019){Sokol}, {Gutermuth}, {Pokhrel}, {G{\'o}mez-Ruiz}, {Wilson}, {Offner}, {Heyer}, {Luna}, {Schloerb}, \& {S{\'a}nchez}}]{sokol2019MNRAS.483..407S}
{Sokol}, A.~D., {Gutermuth}, R.~A., {Pokhrel}, R., {et~al.} 2019, \mnras, 483, 407

\bibitem[{{Solomon} {et~al.}(1987){Solomon}, {Rivolo}, {Barrett}, \& {Yahil}}]{solomon1987}
{Solomon}, P.~M., {Rivolo}, A.~R., {Barrett}, J., \& {Yahil}, A. 1987, \apj, 319, 730

\bibitem[{{Solomon} \& {Wickramasinghe}(1969)}]{solomen1969}
{Solomon}, P.~M. \& {Wickramasinghe}, N.~C. 1969, \apj, 158, 449

\bibitem[{{Verliat} {et~al.}(2022){Verliat}, {Hennebelle}, {Gonz{\'a}lez}, {Lee}, \& {Geen}}]{verliat2022A&A...663A...6V}
{Verliat}, A., {Hennebelle}, P., {Gonz{\'a}lez}, M., {Lee}, Y.-N., \& {Geen}, S. 2022, \aap, 663, A6

\bibitem[{{Wesson} \& {Bevan}(2021)}]{wesson2021ApJ...923..148W}
{Wesson}, R. \& {Bevan}, A. 2021, \apj, 923, 148

\bibitem[{{Williams} \& {McKee}(1997)}]{williams1997ApJ...476..166W}
{Williams}, J.~P. \& {McKee}, C.~F. 1997, \apj, 476, 166

\bibitem[{{Wong} {et~al.}(2022){Wong}, {Oudshoorn}, {Sofovich}, {Green}, {Shah}, {Indebetouw}, {Meixner}, {Hacar}, {Nayak}, {Tokuda}, {Bolatto}, {Chevance}, {De Marchi}, {Fukui}, {Hirschauer}, {Jameson}, {Kalari}, {Lebouteiller}, {Looney}, {Madden}, {Onishi}, {Roman-Duval}, {Rubio}, \& {Tielens}}]{wong2022ApJ...932...47W}
{Wong}, T., {Oudshoorn}, L., {Sofovich}, E., {et~al.} 2022, \apj, 932, 47

\bibitem[{{Wu} {et~al.}(2004){Wu}, {Wei}, {Zhao}, {Shi}, {Yu}, {Qin}, \& {Huang}}]{wu2004A&A...426..503W}
{Wu}, Y., {Wei}, Y., {Zhao}, M., {et~al.} 2004, \aap, 426, 503

\bibitem[{{Xu} {et~al.}(2020{\natexlab{a}}){Xu}, {Offner}, {Gutermuth}, \& {Oort}}]{xu2020ApJ...905..172X}
{Xu}, D., {Offner}, S. S.~R., {Gutermuth}, R., \& {Oort}, C.~V. 2020{\natexlab{a}}, \apj, 905, 172

\bibitem[{{Xu} {et~al.}(2020{\natexlab{b}}){Xu}, {Offner}, {Gutermuth}, \& {Oort}}]{xu2020}
{Xu}, D., {Offner}, S. S.~R., {Gutermuth}, R., \& {Oort}, C.~V. 2020{\natexlab{b}}, \apj, 890, 64

\bibitem[{{Zhang} {et~al.}(2005){Zhang}, {Hunter}, {Brand}, {Sridharan}, {Cesaroni}, {Molinari}, {Wang}, \& {Kramer}}]{zhang2005ApJ...625..864Z}
{Zhang}, Q., {Hunter}, T.~R., {Brand}, J., {et~al.} 2005, \apj, 625, 864

\bibitem[{{Zinnecker} \& {Yorke}(2007)}]{zinnecker2007ARA&A..45..481Z}
{Zinnecker}, H. \& {Yorke}, H.~W. 2007, \araa, 45, 481

\end{thebibliography}
}

\begin{appendix}

\section{Wasserstein Distance}\label{sec: wasserstein distance}

The Wasserstein distance is a similarity metric between two probability distributions. It can be understood as the cost of transport or the work done while converting one distribution to another in the optimal way possible. It is also called the Earth-Mover distance. We use the scipy implementation \url{https://docs.scipy.org/doc/scipy/reference/generated/scipy.stats.wasserstein_distance.html} of the Wasserstein distance to calculate the difference/drift between the distributions of properties in the various feedback bins and report them in the following tables. The drift represents the level of dissimilarity between two distributions and can vary between zero and infinity. We consider that two distributions are dissimilar if they have a drift > 0.1. It is important to note that the Wasserstein distance is only useful for comparing the same properties, e.g. the core radius in different bins. Therefore, the drifts obtained for different properties should not be compared directly.

\begin{table}[]
\caption{Drifts for the distributions of radius (above the diagonal) and mass (below the diagonal) for the global feedback bins obtained using Wasserstein distance.}
\label{tab:wasser_radius_mass_global}
\begin{tabular}{llllll}
\hline
     & All  & No   & Low  & Moderate & High \\ \hline
All  & -    & 0.23 & 0.02 & 0.01     & 0.04 \\
No   & 0.09 & -    & 0.22 & 0.22     & 0.28 \\
Low  & 0.02 & 0.11 & -    & 0.03     & 0.06 \\
Mid  & 0.03 & 0.07 & 0.05 & -        & 0.05 \\
High & 0.02 & 0.11 & 0.02 & 0.05     & -    \\ \hline
\end{tabular}
\tablefoot{We consider two distributions to be different if they have a drift > 0.1.}
\end{table}

\begin{table}[]
\caption{Drifts for the distributions of radius (above the diagonal) and mass (below the diagonal) for the low feedback bins obtained using Wasserstein distance.}
\label{tab:wasser_rad_mass_low}
\begin{tabular}{llll}
\hline
       & All  & Global & Out  \\ \hline
All    & -  & 0.02   & 0.03 \\
Global & 0.02 & -    & 0.01 \\
Out    & 0.02 & 0.04   & -  \\ \hline
\end{tabular}
\end{table}

\begin{table}[]
\caption{Drifts for the distributions of radius (above the diagonal) and mass (below the diagonal) for the moderate feedback bins obtained using Wasserstein distance.}
\label{tab:wasser_rad_mass_mid}
\begin{tabular}{lllll}
\hline
       & All  & Global & Out  & Wind \\ \hline
All    & -    & 0.01   & 0.12 & 0.13 \\
Global & 0.03 & -      & 0.13 & 0.12 \\
Out    & 0.03 & 0.06   & -    & 0.25 \\
Wind   & 0.09 & 0.06   & 0.13 & -    \\ \hline
\end{tabular}
\end{table}

\begin{table}[]
\caption{Drifts for the distributions of radius (above the diagonal) and mass (below the diagonal) for the high feedback bins obtained using Wasserstein distance.}
\label{tab:wasser_rad_mass_high}
\begin{tabular}{llllll}
\hline
       & All  & Global & Out  & Wind & Supernova  \\ \hline
All    & -    & 0.04   & 0.08 & 0.08 & 0.27 \\
Global & 0.02 & -      & 0.04 & 0.12 & 0.32 \\
Out    & 0.14 & 0.12   & -    & 0.16 & 0.35 \\
Wind   & 0.09 & 0.11   & 0.23 & -    & 0.2  \\
Supernova    & 0.26 & 0.28   & 0.39 & 0.17 & -    \\ \hline
\end{tabular}
\end{table}

\begin{table}[]
\caption{Drifts for the distributions of velocity dispersion (above the diagonal) and virial parameter (below the diagonal) for the global feedback bins obtained using Wasserstein distance.}
\label{tab:wasser_radius_mass_global}
\begin{tabular}{llllll}
\hline
     & All  & No   & Low  & Moderate & High \\ \hline
All  & -    & 0.11 & 0.05 & 0.07     & 0.14 \\
No   & 0.15 & -    & 0.07 & 0.18     & 0.25 \\
Low  & 0.09 & 0.08 & -    & 0.11     & 0.18 \\
Mid  & 0.14 & 0.25 & 0.17 & -        & 0.07 \\
High & 0.27 & 0.4  & 0.32 & 0.15     & -    \\ \hline
\end{tabular}
\end{table}

\begin{table}[]
\caption{Drifts for the distributions of velocity dispersion (above the diagonal) and virial parameter (below the diagonal) for the low feedback bins obtained using Wasserstein distance.}
\label{tab:wasser_vel_vir_low}
\begin{tabular}{llll}
\hline
       & All  & Global & Out  \\ \hline
All    & -  & 0.05   & 0.21 \\
Global & 0.09 & -    & 0.17 \\
Out    & 0.41 & 0.37   & -  \\ \hline
\end{tabular}
\end{table}

\begin{table}[]
\caption{Drifts for the distributions of velocity dispersion (above the diagonal) and virial parameter (below the diagonal) for the moderate feedback bins obtained using Wasserstein distance.}
\label{tab:wasser_vel_vir_mid}
\begin{tabular}{lllll}
\hline
       & All  & Global & Out  & Wind \\ \hline
All    & -  & 0.07   & 0.04 & 0.04 \\
Global & 0.14 & -    & 0.11 & 0.06 \\
Out    & 0.16 & 0.29   & -  & 0.06 \\
Wind   & 0.1  & 0.07   & 0.22 & -  \\ \hline
\end{tabular}
\end{table}

\begin{table}[]
\caption{Drifts for the distributions of velocity dispersion (above the diagonal) and virial parameter (below the diagonal) for the high feedback bins obtained using Wasserstein distance.}
\label{tab:wasser_vel_vir_high}
\begin{tabular}{llllll}
\hline
       & All  & Global & Out  & Wind & Supernova  \\ \hline
All    & -  & 0.14   & 0.19 & 0.1  & 0.12 \\
Global & 0.27 & -    & 0.05 & 0.04 & 0.03 \\
Out    & 0.45 & 0.18   & -  & 0.09 & 0.07 \\
Wind   & 0.19 & 0.09   & 0.27 & -  & 0.03 \\
Supernova    & 0.26 & 0.07   & 0.19 & 0.08 & -  \\ \hline
\end{tabular}
\end{table}

\end{appendix}
\end{document}